\documentclass[aps,pra,twocolumn,superscriptaddress]{revtex4-2}
\usepackage{commath}
\usepackage{physics}
\usepackage{amsmath}
\usepackage{mathtools}
\usepackage{amssymb}
\usepackage{comment}
\usepackage{multirow}
\usepackage{color}
\usepackage[svgnames]{xcolor}
\usepackage[colorlinks=true,linkcolor=blue,citecolor=blue,urlcolor=blue]{hyperref}

\usepackage{cleveref}
\crefname{fig}{Fig.}{Figs.}
\crefname{extendedfig}{Extended Data Fig.}{Extended Data Figs.}
\crefname{eq}{Eq.}{Eqs.}
\crefalias{equation}{eq}
\crefname{Methods}{Methods}{Methods Sections}
\crefname{sec}{Section}{Sections}
\crefname{app}{Appendix}{Appendices}
\crefname{table}{Table}{Tables}
\crefname{main}{Main text Section}{Main text Sections}

\newcommand{\eqpair}[4]{%
  \noindent
  \begin{minipage}[b]{#1\linewidth}\begin{equation}#2\end{equation}\end{minipage}%
  \hfill
  \begin{minipage}[b]{#3\linewidth}\begin{equation}#4\end{equation}\end{minipage}%
}

\makeatletter
\def\doauthor#1#2#3{%
  \ignorespaces#1\unskip
  \@if@empty{#2}{}{\frontmatter@footnote{#2}}%
  \@listcomma
  \begingroup #3\endgroup
  \space\@listand
}
\makeatother

\begin{document}

\title[Article Title]{Dynamically modulated light-matter interactions for ancilla-tolerant bosonic control}

\author{Aniket Maiti\textsuperscript{\textcolor{blue}{*,\ddag}}}

\author{John W.O. Garmon}
\thanks{Contributed equally}
\thanks{\\~\protect\raggedright
Corresponding authors:
aniket.maiti@yale.edu,
john.garmon@yale.edu,
robert.schoelkopf@yale.edu}

\author{Daniel K. Weiss}
\altaffiliation{Present address: 
D-Wave Quantum Inc., New Haven, CT 06511, USA}
\author{Takahiro Tsunoda}
\altaffiliation{Present address: Google Quantum AI, Goleta, CA 93111, USA}

\affiliation{Departments of Applied Physics and Physics, Yale University, New Haven, 06511, CT, USA}

\affiliation{Yale Quantum Institute, Yale University, New Haven, 06511, CT, USA}

\author{Yao Lu}
\affiliation{Departments of Applied Physics and Physics,
Yale University, New Haven, 06511, CT, USA}
\affiliation{Yale Quantum Institute,
Yale University, New Haven, 06511, CT, USA}
\affiliation{Superconducting Quantum Materials and Systems
Center, Fermi National Accelerator Laboratory (FNAL),
Batavia, IL 60510, USA}

\author{Steven M. Girvin}
\author{Robert J. Schoelkopf}

\affiliation{Departments of Applied Physics and Physics, Yale University, New Haven, 06511, CT, USA}

\affiliation{Yale Quantum Institute, Yale University, New Haven, 06511, CT, USA}

%%==================================%%
%% Abstract %%
%%==================================%%

\newcommand{\red}[1]{\textcolor{red}{#1}}
\newcommand{\JWOG}[1]{\textcolor{cyan}{#1}}
\newcommand{\DKW}[1]{\textcolor{magenta}{#1}}
\newcommand{\TT}[1]{\textcolor{ForestGreen}{#1}}
\newcommand{\AM}[1]{\textcolor{purple}{#1}}
\newcommand{\YL}[1]{\textcolor{Blue}{#1}}

\begin{abstract}
Strong light-matter interactions are central to state-of-the-art quantum technologies, spanning platforms as diverse as Rydberg atoms, trapped ions, and superconducting circuits. In superconducting circuits, the strong dispersive coupling between microwave photons (light) and ancillary qubits (matter) has enabled universal bosonic control and quantum error correction.
However, this always-on dispersive interaction propagates errors from the ancillary qubits to the bosonic systems both while idling and when performing operations, introducing a trade-off between control strength and fidelity. 
We break this trade-off by introducing a parametrically activated hybridization between the bosonic modes and ancilla qubits, keeping both elements decoupled when not active.
When activated, this hybridization mediates a dynamic light-matter coupling with non-trivial engineered symmetries, permitting resilience to ancilla decay and dephasing.
This dynamic coupling functions as a new control primitive that offers a route to bosonic processors whose performance is set by the oscillator, rather than the ancilla.
\end{abstract}

% \keywords{dynamic coupling, fault-tolerant control}

\maketitle

\section*{Introduction}
Light-matter interactions form a central pillar of the study of quantum physics today.
The strength of these interactions dictates both the speed of quantum control, and the rate of decoherence of quantum systems.
Modern light-matter experiments enhance this interaction strength by trapping light in a standing wave, for example in an optical or microwave cavity with Rydberg atoms~\cite{Rempe_cavity_QED,haroche2006exploring}, or in a superconducting microwave resonator coupled to an artificial atom~\cite{wallraff2004strong,schoelkopf2008wiring}.
The resulting off-resonant Jaynes-Cummings interaction hybridizes the light and atom(s), producing dressed modes that have a strong and always-on nonlinear interaction.
This effect has served as a fundamental building block in modern superconducting architectures, enabling high-fidelity qubit readout \cite{WalraffDispersiveReadout2005,WalraffDispersiveReadout2017,Swiadek2023,WeiSumeruReadout} and entangling gates \cite{Majer2007,DiCarlo2009,PaikRIPGate2016}, and conversely, allowing qubits to manipulate the oscillator.
We focus on the latter, which includes the preparation, control, and tomography of highly non-classical states in a microwave oscillator with ancillary qubits~\cite{Brune1992CatState,Bertet2002Fock1Wigner,Schuster2007PhotonNumberSplitting,Hofheinz2009,VlastakisBigCat2013,Krastanov2015,SNAP_2015,Wang2016CatTwoBoxes}.

\begin{figure*}[ht]
\centering
\includegraphics[width=0.98 \textwidth]{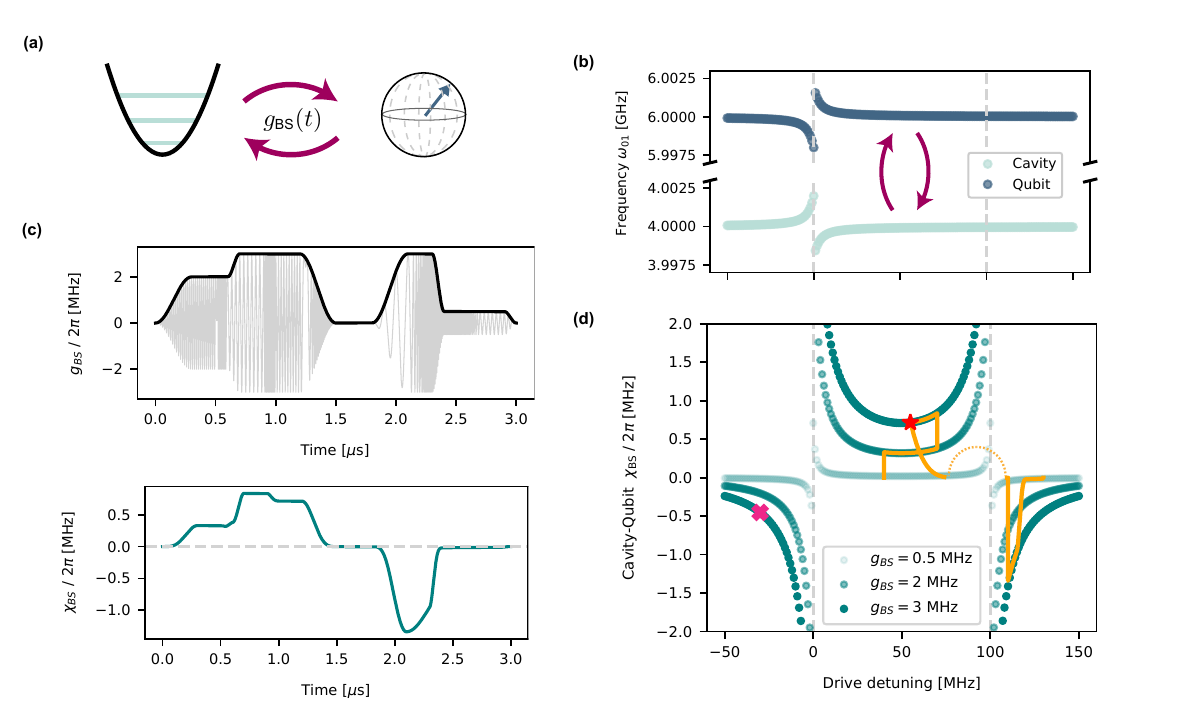}
\caption{
\textbf{Dynamic modulation of qubit-cavity dispersive shift}
\textbf{a}, Illustration of qubit-cavity exchange interaction, dynamically activated by a parametric coupler (not shown).
\textbf{b}, Numerically simulated avoided crossing between the qubit and cavity, with a realistic parametric coupling strength of $g_{\text{BS}} = 3$ MHz. The coupling is orders of magnitude smaller than the static detuning of the two modes, yet it can activate dispersive shifts of the same order of magnitude.
\textbf{c}, Modulating the parametric coupling (top panel, pulse shown in gray with overall amplitude outlined in black) results in a change of the effective qubit-cavity coupling strength and detuning. The resulting dispersive shift for a transmon qubit can change between positive and negative values within a single control pulse (bottom panel).
\textbf{d}, Dispersive shifts (teal) as a function of coupling strength and detuning between a cavity and a transmon with Kerr $\alpha=100$ MHz. The modulated coupling in \textbf{c} produces the dynamic trajectory shown in yellow, which includes the avoidance of the pole at $\Delta_{\text{BS}}=\alpha$. Marked points include the activation of a typical hybridization ($g_{\text{BS}}/\Delta_{\text{BS}} = 0.1$, pink cross) to get $\chi_{\text{BS}}/2\pi = 0.5$ MHz, and a special point where the dispersive shift is exceptionally linear ($\chi'_e=0$, red star).
}\label[fig]{fig:Fig1}
\end{figure*}

While strong cooperativity of the oscillator and the ancilla is beneficial for control, an always-on interaction can be harmful.
In particular, when the coherence of the oscillator and the ancilla differ significantly, errors in one mode can propagate and spoil the quantum information stored in the other.
This issue is especially visible in superconducting circuits, where many demonstrations of bosonic control are limited by errors in the higher-loss ancillary circuit \cite{BeyondBreakeven2016,GKPExperiment, PhilSergeFTDetection, Reinhold2020}.
Indeed, the coherence of the ancilla has been directly shown to limit state-of-the-art error-correction performance \cite{GKPExperiment,sivak_real-time_2023}.
In qubit-based superconducting experiments,tunable couplers have been utilized to shield against ancilla interactions when they are undesired~\cite{FeiYanCouplerTheory2018,SungTunableCouplerExperiment2021}.
However, these tunable couplers do not immediately work in the bosonic setting, as they often introduce the same nonlinear error channels to the bosonic mode as the ancilla qubit does.
We propose a novel way to break this trade-off by dynamically modulating the oscillator-ancilla hybridization through parametrically driven interactions.
This allows the activation of full dispersive control only when it is desired, and provides additional knobs to create symmetries in the interaction that enable a degree of native fault-tolerance.

Previous experiments in superconducting circuits have demonstrated in-situ modulation of the hybridization between an oscillator and a transmon qubit by changing the ancilla's frequency using fast-flux control \cite{DiCarlo2009,Martinis2014, Fei2018,Swiadek2023,yvonne_fluxTuning_2024}.
This technique only permits tuning between effectively static couplings, and one must tune continuously between these static values and thus pass through any intermediate unwanted poles in the interaction.
An alternative strategy that has shown some success is utilizing parametrically driven interactions, where an off-resonant drive on the transmon tunes its effective coupling with the oscillator~\cite{Reinhold2020,PhilSergeFTDetection,Akshay_chi_tuning,ParametricCQEDPaper}.
In particular, previous work demonstrated a matching of the dispersive shifts of the first- and second-excited transmon states.
This created a symmetry \cite{KapitErrorTransparency2018} that prevented the propagation of transmon errors to the oscillator mode.
However, due to the static hybridization of the transmon and oscillator, this engineered interaction was oscillator state-dependent.
Moreover, stronger drives on the transmon inevitably reduced its coherence, thus limiting the possible applications of this technique.

In this work, we introduce novel control techniques that dynamically dress the oscillator and qubit using a parametric beamsplitting interaction.
Such techniques are enabled by recent progress with purpose-built parametric couplers~\cite{ChapmanSNAILBeamsplitterPaper,MaitiLINCTheory2025}, which activate strong exchange interactions when driven by an RF drive, while minimizing unwanted static nonlinearities and coupler excitation.
Importantly, the strength and effective detuning of this parametric exchange interaction can be set to high precision by the drive, allowing one to cherry-pick symmetry points in the interaction.
The general regime of `parametric circuit-QED' has been previously proposed as beneficial in the qubit context, with initial demonstrations of qubit readout and transmon gates \cite{ParametricCQEDTheory,ParametricCQEDPaper,Jin2024ParametricCZNIST}.
We expand such interactions to encompass near-arbitrary modulation of an oscillator-qubit coupling, allowing a new control architecture for bosonic quantum processors \cite{LiuISA2026}.

The remainder of the paper is organized as follows.
In \cref{sec:sec1}, we show the landscape of the effective driven nonlinear interaction between the qubit and oscillator over a wide range of coupling strengths and detunings.
This effective non-linear landscape only appears in the adiabatically dressed frame, but we show that we can modulate through this dressed frame within the timescale of a single gate operation.
 The flexibility of this technique allows us to circumvent unwanted poles in the interaction, even allowing positive and negative dispersive shifts within a single control pulse.
In \cref{sec:echoed}, we illustrate the benefits of such a time-dependent interaction.
First, we show that it can dynamically decouple the oscillator from low-frequency noise affecting the ancilla.
Second, we show that a bi-chromatic coupling, enabled by driving the parametric coupler at two tones simultaneously, activates effectively static interactions that are beyond the physics allowed by any true static coupling.
This allows one to turn on the beneficial symmetry proposed in \cite{KapitErrorTransparency2018,PhilSergeFTDetection,Reinhold2020}, while avoiding its pitfalls.
Finally, we show numerically optimized pulses that directly optimize within the full landscape of the dynamic dispersive interaction, enabling novel ways to measure error syndromes of a bosonic logical qubit while remaining first-order insensitive to both ancilla loss and low-frequency dephasing.
Overall, this control technique allows natively "ancilla-tolerant" bosonic architectures, both while idle and when activating entangling gates.

\section{Dynamic Oscillator-Ancilla interaction}
\label[sec]{sec:sec1}
\subsection{Regimes of interaction}
\label[sec]{sec:regimes}

The use of a parametric coupling to control the oscillator-ancilla interaction permits the investigation of previously under-explored coupling regimes.
Specifically, we consider the interaction Hamiltonian
\begin{align}
    H_0/\hbar &= H_{\rm{NL}}(\hat{q},\hat{q}^\dagger)/\hbar +\Delta_{s} \hat{q}^\dagger \hat{q} \label{eq:interaction_Hamiltonian}\\ \nonumber &\quad+ g_{\rm BS}(t)e^{i\int_0^t\omega_d(\tau)d\tau}\hat{a}^\dagger\hat{q} + g_{\rm BS}^{*}(t)e^{-i\int_0^t\omega_d(\tau)d\tau}\hat{a}\hat{q}^{\dagger},
\end{align}
where the parametric interaction is actuated by a drive with time-dependent strength $g_{\rm BS}(t)$ and frequency $\omega_d(t)$, see \cref{fig:Fig1}(a-c). The oscillator ($\hat{a}$) and ancilla ($\hat{q}$) have a static detuning $\Delta_s=\omega_q-\omega_a$, and the ancilla nonlinearity is denoted by $H_{\rm{NL}}$.
To eliminate the oscillatory time dependence of the beamsplitter interaction in favor of an explicit detuning, we perform a rotating-frame transformation defined by the unitary $U_{\rm{rot}} = \exp(i\int_0^t\omega_d(\tau)d\tau~q^\dagger q)$ , yielding
\begin{align}
\label{eq:parametric_detuned_bs}
    H_{p}/\hbar
&= H_{\rm{NL}}(\hat{q},\hat{q}^\dagger)/\hbar+\Delta_{\rm BS}(t) \hat{q}^\dagger \hat{q} \\
&+ g_{\rm BS}(t)\hat{a}^\dagger\hat{q} + g^*_{\rm BS}(t)\hat{a}\hat{q}^\dagger.\nonumber
\end{align}
We have defined the time-dependent drive detuning $\Delta_{\rm{BS}}(t) = \Delta_s - \omega_d(t)$, while the complex coupling $g_{\rm BS}(t)$ retains independent control of the interaction amplitude and phase.
Notice that this Hamiltonian closely resembles that of the generalized Jaynes-Cummings model \cite{JensTransmonPaper, circuitQEDReview} 
\begin{align}\label{eq:detuned_bs_w_NL}
    H_{s}/\hbar = H_{\rm NL}(\hat{q},\hat{q}^\dagger)/\hbar +\Delta \hat{q}^\dagger \hat{q} + g(\hat{a}^\dagger\hat{q} + \hat{a}\hat{q}^\dagger ), 
\end{align}
where here the coupling strength and detuning are set by the static mode frequencies and dipole-dipole coupling strengths. 
In the parametric case, 
the drive supplies the energy necessary to bridge the gap between the two mode frequencies, hybridizing the oscillator with the nonlinear ancilla only when necessary.
When the drive is off, the oscillator and ancilla are decoupled, leading to a large on-off ratio.

Both Hamiltonians $H_{p}$ and $H_{s}$ can give rise to state-dependent frequency shifts, such as the cross-Kerr interaction between a transmon and a cavity \cite{Schuster2007PhotonNumberSplitting,JensTransmonPaper, circuitQEDReview,ParametricCQEDPaper,ParametricCQEDTheory}.
In fact, \cref{eq:parametric_detuned_bs,eq:detuned_bs_w_NL} only differ due to the time dependence of $g_{\rm{BS}}$ and $\Delta_{\rm{BS}}$. For now, we consider the case where a single-tone coupling is adiabatically turned on to a steady-state value, deferring the treatment of finite ramps to \cref{sec:speed_limit}.
Below, we derive perturbative quantities in terms of $g, \Delta$ and the strength of the ancilla's nonlinearity $K$.
For the remainder of the paper, we assume the nonlinearity to be a Kerr oscillator with $H_{\rm NL}/\hbar = -\frac{K}{2} \hat{q}^\dagger\hat{q}^\dagger \hat{q} \hat{q}$.

\begin{figure}[ht]
\centering
\includegraphics[width=0.48 \textwidth]{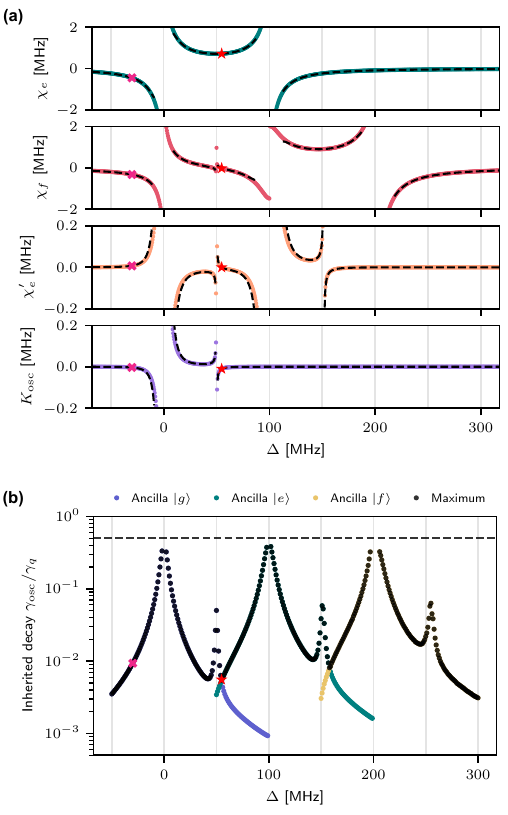}
\caption{
\textbf{Effects of dynamic dressing}
\textbf{a}, Coherent nonlinearities activated by the parametric dressing, as a function of detuning from resonance with $\omega_{ge}$, extracted from direct diagonalization of \cref{eq:detuned_bs_w_NL} (solid lines) along with analytic predictions from \cref{app:PertTheory} (dashed lines). Dispersive shifts to different transmon levels $\chi_e$ (teal) and $\chi_f$ (red), the nonlinearity of the oscillator-ancilla dispersive shift ($\chi_e'$, peach), and the inherited cavity nonlinearity ($K_{\rm{osc}}$, purple) all exhibit multi-pole features and the changing of sign. 
Here, $g$ and $K$ are set to be $3$ MHz and $100$ MHz, respectively.
\textbf{b}, Inherited decay of the oscillator extracted from time-domain simulations (see text), expressed in units of the transmon decay rate. 
The inherited decay shows peaks at the various poles, with the decoherence saturating to $\gamma_q/2$ at maximal hybridization.
Special operating points are marked in pink ($g/\Delta\sim 0.1$) and red ($\chi'=0$).
}\label[fig]{fig:Fig2}
\end{figure}

The general structure of these shifts depends on the hierarchy of $g,\Delta,$ and $K$.
Typical circuit-QED experiments work in the regime $g \sim K \ll \Delta$, which yields a dispersive interaction robust to fabrication variations while still enabling universal control~\cite{Krastanov2015,SNAP_2015}. 
Interestingly, while a parametric $g_{\rm{BS}}$ is generally smaller than a static dipole-dipole $g$, by also tuning
$\Delta_{\rm{BS}}$ to operate in the regime $g < \Delta \sim K$, we predict dispersive-interaction strengths comparable to the static case. 
Indeed, as indicated by the pink crosses in \cref{fig:Fig1}d and \ref{fig:Fig2}, dispersive strengths approaching MHz scale are attainable with $g/\Delta = 0.1$.
Further, we can access in situ the finer structure of the dispersive interaction, as described below.

Although motivated here by a parametrically activated interaction, the results in this section depend only on the presence of a detuned beamsplitter coupling between a linear oscillator and a Kerr-nonlinear ancilla, regardless of implementation.
We derive the state-dependent shifts by treating the ancilla-oscillator interaction of \cref{eq:detuned_bs_w_NL} perturbatively.
We proceed up to fourth order in perturbation theory, where many relevant quantities (such as inherited oscillator Kerr) first appear. The effective Hamiltonian is given by 
\begin{align} \label{eq:dispersive_H_w_NL}
    H_{int}/\hbar = \tilde{\Delta} \hat{q}^\dagger \hat{q}
    &+\left(\chi_e + \frac{\chi_e'}{2}[\hat{a}^\dagger \hat{a}-1]\right)\hat{a}^\dagger \hat{a}\ket{e}\bra{e} \\ \nonumber
    \enspace &+\left(\chi_f + \frac{\chi_f'}{2}[\hat{a}^\dagger \hat{a}-1]\right)\hat{a}^\dagger \hat{a}\ket{f}\bra{f} \nonumber \\
    \enspace 
    &-\frac{K}{2} \hat{q}^\dagger\hat{q}^\dagger \hat{q} \hat{q}
    +\frac{K_{\rm osc}}{2}\hat{a}^\dagger\hat{a}^\dagger \hat{a} \hat{a} +O([g/\Delta]^6),\nonumber
\end{align}
see \cref{app:PertTheory} for details). We have defined the Lamb-shifted detuning $\tilde{\Delta}=  \Delta + g^2/\Delta$, and only consider excitations up to the $\ket{f}$ level of the ancilla.
The nonlinear interaction strengths are given to leading order in terms of $g,\Delta$ and $K$ as 
\begin{align}
    \chi_e &=     -\frac{2g^2 K }{\Delta (\Delta- K)} 
        % +\frac{4g^4K(K^2 + 2\Delta(\Delta-K))}{\Delta^3(\Delta-K)^3}, 
        \\
    \chi_f &= - \frac{2 g^2 K(2\Delta -K)}{\Delta(\Delta-K)(\Delta -2K )} \\
    % \chi_e' &=      \frac{4g^4K^2(-3K^3 + 11K^2\Delta - 15K\Delta^2 + 9\Delta^3)}
    %     {\Delta^3(K - 2\Delta)(3K - 2\Delta)(K - \Delta)^3}, \\
    K_{\rm osc} &=     -\frac{2g^4K}{\Delta^3(2\Delta- K)}.
\end{align}
Expressions for $\chi_e'$ and $\chi_f'$, as well as the full fourth-order Hamiltonian are given in \cref{app:PertTheory}.
The value of $\chi_e$ changes sign when traversing across the poles at $\Delta$ and $\Delta-K$, making $\chi_e$ positive in the so-called ``straddling regime", see \cref{fig:Fig1}(d) \cite{JensTransmonPaper}. 
Additionally, we find poles not present in the dispersive shift $\chi_e$ that appear in the inherited oscillator Kerr $K_{\rm osc}$ and in the nonlinear correction to the dispersive shift $\chi_e'$, revealing the rich structure displayed in \cref{fig:Fig2}(a).
While here we assume a transmon spectrum and its matrix elements, the perturbative framework is quite general, and could be applied to other nonlinear ancillae such as fluxonium \cite{Zhu2013Guanyu4thOrder,KeNieFluxoniumOscillator2026,ShoumikShantanuFluxoniumGKP2026Arxiv}.

Along with the coherent nonlinear effects present in \cref{eq:dispersive_H_w_NL}, the oscillator also inherits 
decoherence from the ancilla,
as shown in~\cref{fig:Fig2}(b). 
To quantify the inherited oscillator decay, we perform time-domain simulations in which the hybridization is activated adiabatically, varying the length of the interaction.
We initialize the transmon in either the $\ket{g}$, $\ket{e}$, or $\ket{f}$ state, subject to single-photon loss. 
In each case, we track the no-jump evolution of an oscillator coherent state, so that the transmon remains in its prepared state and the extracted rate reflects the oscillator's inherited decay alone.
The maximum of these rates across the three transmon states is shown in \cref{fig:Fig2}b.
The oscillator experiences an enhanced decay rate when the oscillator frequency $\omega_{a}$ is near-resonant with one of the ancilla transitions $\omega_q^{n,n+1}$, as predicted by the Purcell effect \cite{Purcell1946}.
This decoherence peaks at $\gamma_q/2$, as expected when the modes are maximally hybridized (and $g \gg \gamma_q)$.
% \YL{this is only true when the decay rate is much lower than the coupling strength} 
Additionally, when a two-photon transition in the ancilla becomes resonant with twice the oscillator frequency,
$2\omega_{a} \approx \omega_q^{n,n+2}$, the oscillator suffers from two-photon loss, leading to the sharper peaks in \cref{fig:Fig2}b.
The hybridization can also introduce oscillator dephasing, either through the dispersive shift with a thermally excited ancilla, or by directly inheriting frequency noise from the ancilla.

Having established the coherent and incoherent effects of the hybridization, we now consider how they constrain the choice of parametric drive parameters.
% $g_{\rm BS}$ and $\Delta_{\rm BS}$.
To generate the strongest possible coupling to the ancilla for fast operations, one reasonable approach could be to drive precisely on resonance and completely hybridize the two modes.
While this method certainly has applications like state preparation and tomography \cite{Law1996ResonantControl,Hofheinz2009,yvonne_fluxTuning_2024,Huang2025fastsidebandcontrolweakly,huang2026JCcontrol}, it also fully exposes the oscillator to the decoherence of the ancilla.
Instead, we can operate at a small but finite detuning, with $\Delta_{\rm{BS}}$ roughly an order of magnitude larger than the coupling strength $g_{\rm BS}$, resulting in a hybridization similar to usual static regimes.
Note that the oscillator will still inevitably inherit decoherence from the mediating parametric coupler, motivating the use of couplers that are minimally invasive \cite{SQUIDBeamsplitterPaper,ChapmanSNAILBeamsplitterPaper,MaitiLINCTheory2025,Iivari2024}.
We explore effects specific to the coupler implementation in more detail in \cref{app:coupler}. 

Vitally, while the landscape is more complex in the $g < \Delta \sim K$ regime, we have high precision over our choice of $g_{\rm{BS}}$ and $\Delta_{\rm{BS}}$ as they are controlled by room-temperature RF electronics.
This enables fine-tuning the relative effects of various poles, allowing bosonic control at particularly desirable operating points. 
For instance, we can target the point $\chi_e' = 0$ (marked in Figs. \ref{fig:Fig1}d and \ref{fig:Fig2}b by the red star) where the dispersive shift becomes exceptionally linear.
Access to such points permits bosonic encodings that may impose stringent limits on inherited nonlinearity \cite{GKPExperiment,sivak_real-time_2023}.

\subsection{Speed limits on dynamic hybridization}
\label[sec]{sec:speed_limit}

\begin{figure}[ht]
\centering
\includegraphics[width=0.48 \textwidth]{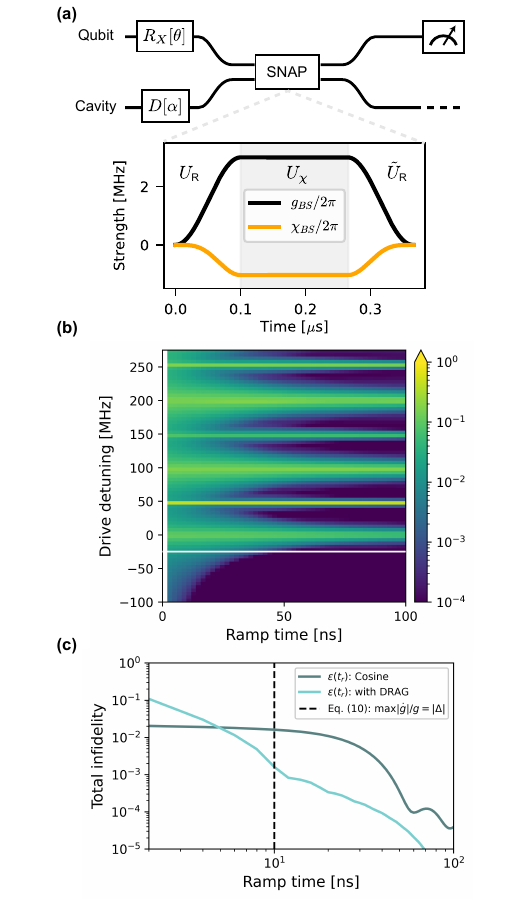}
\caption{
\textbf{Speed limits on dynamic dressing}
\textbf{a}, Example pulse with cosine ramps, where the dynamic dressing is activated (shaded) selectively during an entangling Unitary $U_\chi$. We analyze the errors during the ramps $U_R$.
\textbf{b}, Total unitary infidelity during ramps, as a function of ramp time and detuning from poles of interaction.
There are poles visible at detunings of multiples ancilla anharmonicity K/2, and these contribute through different levels of the ancilla.
\textbf{c}, Total infidelity (green) as a function of ramp time at $\Delta_{\text{BS}}=-25$ MHz, cross section at white line in \textbf{b}. Applying a DRAG correction improves this infidelity (blue).
}\label[fig]{fig:Fig3}
\end{figure}

In the previous section, we showed that we can specify $\chi$ in situ over a broad range by tuning the linear coupling strength $g_{BS}$ and detuning $\Delta_{BS}$.
We now examine how quickly this interaction can be activated. Rapid ramping may introduce coherent errors, such as unwanted population swap between the bare and the dressed bases, which leads to leakage outside of the bosonic code space.
This prevents us from faithfully moving between unselective operations and selective entangling operations (\cref{fig:Fig3}a).
We demonstrate that the ramp time is ultimately limited by the detuning $\Delta$ from the nearest pole.
Unlike a standard two-level system, the transmon's finite anharmonicity introduces multi-pole characteristics that complicate the adiabatic mapping between the bare and dressed bases. 

We now present a detailed analysis of these non-idealities, followed by concrete ramp protocols inspired by the DRAG formalism \cite{motzoi2009} to minimize leakage near the adiabatic limit.
To find an analytic estimate of the speed limit, 
we model the time-dependent interaction as a hybridization parametrized by the mixing angle $\theta(t)=\frac{1}{2}\arctan\left(\frac{2g_{\rm{BS}}(t)}{\Delta_{\rm{BS}}(t)}\right)$ due to a time-varying coupling $g_{\rm{BS}}(t)$ and detuning $\Delta_{\rm{BS}}(t)$.
By iteratively performing the unitary transformation \cite{motzoi2009}
\begin{align}\label{eq:transformation}
    V_0(t)=\exp(-\theta(t)(\hat{a}^{\dagger}\hat{q}-\hat{a}\hat{q^{\dagger}})),
\end{align}

from the bare basis to the dressed basis via
$H_{\rm{dressed}}= V_0H_{\rm{bare}}V_0^{\dagger} + i\hbar\dot{V}_0V_0^{\dagger}$, we track the effect of the inertial term $i\hbar\dot{V}_0V_0^{\dagger}$ which is the origin of coherent errors. 

We can define the analytic speed limit for adiabatic transitions by requiring the magnitude of the inertial term to be much smaller than the energy-level splitting induced by the transition, 
\begin{align}
\label{eq:adiabatic_condition}
|i\hbar\dot{V}_0V_0^{\dagger}|=|\dot{\theta}(t)|\ll\sqrt{(2g_{\rm{BS}}(t))^2 + \Delta_{\rm{BS}}(t)^2}.
\end{align}
Here, we simplify our analysis by focusing on the scenario in which the detuning $\Delta_{\rm{BS}}$ is fixed, and only the coupling amplitude $g_{\rm{BS}}(t)$ is ramped. 
Under this simplification, and in the regime of $g \ll \Delta_{\rm{BS}}$, we define the speed limit as,
\begin{align}
\rm{max}\{\dot{\it{g}}\}/\it{g} = \rm{\Delta}_{\rm{BS}}
\end{align}
as shown in the black dashed vertical line in \cref{fig:Fig3}c (see \cref{app:speed_limit} for the derivation).
Staying below this speed limit while also remaining in the dispersive regime ensures that \cref{eq:adiabatic_condition} is satisfied.

While this analysis gives a rough estimate of the energy scales involved for proper dynamic hybridization, it fails to capture the full picture. 
So far, for simplicity, we have modeled one of the modes as a two-level system to analyze its energy splitting and derive the adiabatic condition. However, in a typical Circuit QED setup, the ancilla is a weakly anharmonic transmon qubit. The presence of its higher energy levels introduces multiple detuning points that creates several energy splittings within the bandwidth of interest. 
This makes it difficult to satisfy the adiabaticity condition, as the relevant detuning is no longer $\Delta_{\rm{BS}}$ explicitly, but must include the effect of multiple poles (bright region in \cref{fig:Fig3}b).
Although each pole provides a contribution to the total infidelity, the adiabatic condition derived above captures the speed limit as long as the rate of change of the coupling $g_{\rm{BS}}$ is small compared to the magnitude of the detuning from the closest pole, and $\Delta_{\rm{BS}}$ is redefined to be the detuning from this pole.

To quantify the speed limits of dynamically activated dispersive interaction, we evaluate the process infidelity of the ramp-up propagator $\mathcal{U}(t_{\mathrm{r}},0)$ relative to an ideal target unitary $U_{\mathrm{target}}$ that maps each bare product state to the corresponding dressed eigenstate with the correct dynamical phase,
\begin{eqnarray}
    \varepsilon(t_{\rm{r}})=
    1-
    \frac{1}{d^{2}}
    \Bigl|
        \operatorname{Tr}\!\bigl[
        U_{\mathrm{target}}^{\dagger}(t_{\mathrm{r}})\
        \mathcal{U}(t_{\mathrm{r}}, 0)
        \bigr]
    \Bigr|^{2},
\end{eqnarray}
where $d=N_qN_c$ is the dimension of the truncated transmon-cavity Hilbert space used in the simulation. 
$t_r$ is the time at which the pulse ramps up completely.
This metric is sensitive to both population leakage and coherent phase errors, and weighs all levels in both the oscillator and ancilla equally.

We conduct numerical simulations of a time-dependent drive with ramp-up times varying from $t_{\rm{r}}=$2~ns to 100~ns using the Schr\"odinger equation solver in \texttt{dynamiqs} \cite{dynamiqs}. 
For these simulations, we employ a detuned beamsplitter pulse with a cosine ramp followed by a flat top (\cref{fig:Fig3}a); this is a typical configuration for beamsplitter pulses (see, for instance, Ref.~\cite{degraaf2025}) that implements an effectively static drive-induced $\chi$.
This sensitivity to non-adiabaticity is further amplified when operating in the straddling regime between multiple poles.
As shown in \cref{fig:Fig3}c, the infidelity is $\varepsilon < 10^{-4}$ for ramp times exceeding 78~ns for $\Delta_{\rm{BS}}=-25 \rm{MHz}$ (white dashed line in \cref{fig:Fig3}b), whereas shortening the ramp leads to a steep increase in infidelity.
As discussed earlier, this non-ideality arises from the inertial term, which corresponds to the non-adiabaticity of the ramp. 
For shorter ramp times, this inertial term becomes increasingly significant, and its cumulative impact throughout the ramp leads to the observed infidelity.

To gain a more complete picture of the inertial term's effect, we examine the system's performance over a wider parameter space. 
Specifically, we sweep the detuning of the beamsplitter drive, $\Delta_{\rm{BS}}$, in addition to the ramp time. The resulting two-dimensional plot of the gate error reveals multiple sharp features, or poles, where the performance is significantly degraded [\cref{fig:Fig3}(b)]. These poles correspond to resonant swap interactions between the ancilla and a bosonic mode; hence, we want to operate sufficiently away from these poles.
The appearance of multiple poles highlights the multi-level nature of the ancilla, which introduces more complex physics than that described by the two-level system approximation. 
To clarify the role of the interaction poles visible in~\cref{fig:Fig3}, we examine how the population mapping error depends on the initial ancilla state in \cref{app:ancilla-init}.

Finally, we mitigate the effect of the inertial term using a pulse modulation technique known as Derivative Removal by Adiabatic Gate (DRAG).
This technique is widely used to suppress leakage errors in various hardware platforms \cite{motzoi2009, theis2018}.
Although at first glance, the frame transformation (\cref{eq:transformation}) seems significantly different
from the original DRAG scheme \cite{motzoi2009}, 
we can implement a straightforward adaptation of the protocol because the beamsplitter interaction has the same SU(2) symmetry as qubit control \cite{tsunoda2023}.
To suppress the dominant error arising from the inertial term, we focus on mitigating the leakage induced by the nearest pole of interest.
As illustrated in \cref{fig:Fig3}b and \cref{fig:Fig3}c, the application of DRAG enables a reduction of the ramp-up time to as low as 39~ns.
Therefore, we can observe that modest pulse shaping lead to faster activation of the dynamic dispersive interaction.
Lastly, although the standard DRAG is only a partial remedy, a more advanced remedy addressing the multiple poles would likely do the job \cite{theis2018,YaxingLeakageCancellation2025},
but is left for future work.

\section{Dynamic hybridization for useful control}
\label[sec]{sec:echoed}

We now show that parametric hybridization can allow dynamic or effective interactions that provide functionality beyond the reach of static hybridizations.
We focus on bosonic control and error syndrome measurement techniques enabled by such dynamics.

\begin{figure*}[ht]
\centering
\includegraphics[width=0.98 \textwidth]{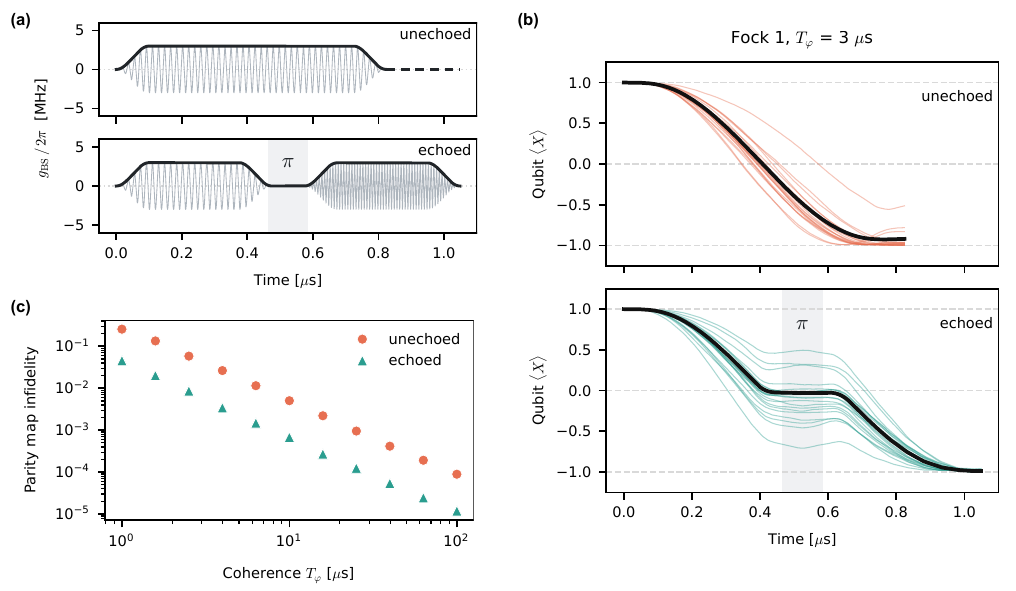}
\caption{
\textbf{Dynamically decoupled hybridization for parity measurements}
\textbf{a}, Calibrated pulses for un-echoed and echoed parity map sequences. The former  turns on a positive dispersive shift ($g_\text{BS}/2\pi\sim3$ MHz, $\Delta_\text{BS}\sim 54.8$ MHz) for a duration such that $\int\chi(t)dt=\pi$. The latter turns on two separate pulses at $g_\text{BS}/2\pi\sim3$ MHz, $\Delta_\text{BS}\sim 54.8, 120$ MHz, such that $\int \chi_{1,2}(t) dt = \pm \pi/2$, respectively, with an echo on the qubit in between.
  \textbf{b}, Monte Carlo simulations for individual qubit trajectories with low-frequency dephasing noise when the oscillator is in Fock $|1\rangle$, plotted in the Lamb-shifted frame for visualization. Colored lines show different trajectory samples, and the black line represents their average.
  The echoed pulse demonstrates refocusing.
  The strength of low-frequency noise here is set such that the transmon has Ramsey $T_\varphi = 3 \mu$s.
  \textbf{c}, Parity map infidelity, averaged over cavity states $\ket{0}$ and $\ket{1}$, as a function of effective transmon Ramsey coherence $T_\varphi$.
}\label[fig]{fig:Fig4}
\end{figure*}

As a first example, we introduce the ability to circumvent poles in the oscillator-ancilla interaction mid-control pulse.
The rapid change in both amplitude and sign of the nonlinear interactions in~\cref{eq:dispersive_H_w_NL} across these poles offers opportunities for dynamical decoupling and noise suppression in bosonic control pulses.
In a traditional light-matter setting, one would either operate at a fixed point far from these poles~\cite{VlastakisBigCat2013,Reagor2016QuantumMemory,BeyondBreakeven2016}, or use frequency tunability to approach these poles when required~\cite{Hofheinz2009,yvonne_fluxTuning_2024}.
This means that typically, one is restricted to exploring the interaction space in only one-dimensional space of the oscillator-qubit detuning $\Delta$, encountering Landau-Zener-like effects when attempting to cross an interaction pole.
Parametric couplings generalize such interactions to the two-dimensional control space of both $\{g, \Delta\}$, allowing one to move across 
such poles with no change in the quantum state.

To demonstrate how this is useful in a control scheme, consider an oscillator-ancilla system where the ancilla experiences low-frequency noise.
In circuit-QED, such noise could arise from $1/f$ flux-noise (e.g., from a parametric coupler), a fluctuating two-level system in the ancilla's environment, or other forms of inhomogeneous broadening.
A well-established strategy to suppress such low-frequency noise is to use a dynamic decoupling pulse sequence on the ancilla \cite{Bylander2011,FeiYan2013}.
However, such an echo on the ancilla also exactly echoes out 
any static dispersive shifts it might have with the oscillator.
For instance, if the interaction Hamiltonian is $H_{\rm I}=\chi\sigma_{z}a^{\dagger}a$, echoing the qubit ($\hat{\sigma}_z \rightarrow -\hat{\sigma}_z$) negates the effects of this interaction, forbidding its use in static bosonic control.
With dynamic hybridization, one can split any period of dispersive interaction into two equal halves, such that each half has the same magnitude, but opposite sign of the dispersive shift ($\chi \rightarrow -\chi$).
If the qubit is echoed in between these two halves, when $g=0$, the pulse preserves the phase accumulated from the dispersive interaction while echoing out all other low-frequency fluctuations on the qubit.
As an additional bonus, there is no selectivity condition on the qubit echo pulse, since the coupling is turned off, while in a static-coupled system the same pulse would require a wide bandwidth dependent on the cavity population.

We demonstrate this pulse by simulating a parity measurement of the oscillator state through a transmon ancilla.
This operation is necessary for detecting and correcting errors in rotation-symmetric bosonic codes, where photon-number parity is a good error syndrome~\cite{mirrahimi_cat_2014,Michael2016,Arne_rotating_2020}, yet is typically limited by ancilla decoherence.
As a point of comparison, we first implement an un-echoed version by activating a detuned coupling at $\Delta_{\text{BS}}/2\pi=54.8$ MHz (the operating point where $\chi'_e=0$) and $g_\text{BS}/2\pi=3$ MHz, with $100$ ns cosine ramps, and finely calibrate the flat-top time to implement a $+\pi$ phase shift per photon.
In the presence of low-frequency dephasing noise on the transmon, individual trajectories spread out significantly (see~\cref{fig:Fig4}b), leading to a low-fidelity pulse.
We then separate the pulse into two halves, with the first remaining at $\Delta_{\text{BS}}/2\pi=54.8$ MHz, but the second half driven at $\Delta_{\text{BS}}/2\pi=120$ MHz to provide an opposite sign of the dispersive shift (\cref{fig:Fig4}a).
The echoed pulse refocuses trajectories and offers a significant infidelity improvement (see \cref{fig:Fig4}c), likely eliminating this error mechanism in practice for any coherence time above $10 \mu s$.
Details for simulating low-frequency noise are provided in~\cref{app:low_freq_echo}.
A similar echo technique may be optimized to cancel effects of other coherent terms, like inherited oscillator Kerr.

\begin{figure*}[ht]
\centering
\includegraphics[width=0.98 \textwidth]{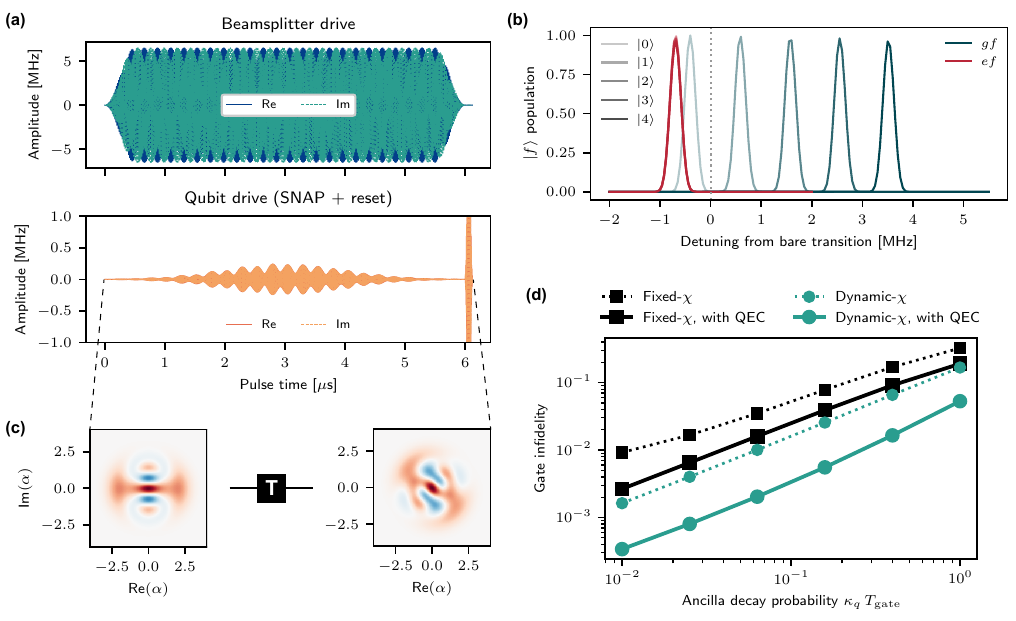}
\caption{
\textbf{Ancilla decay-tolerant gates with a bichromatic interaction}
\textbf{a}, Calibrated pulses for a bichromatic (two-tone) beamsplitter drive and qubit-selective SNAP pulses, along with a final reset $\pi$-pulse, that performs a logical T gate on the kitten code.
  \textbf{b}, The beamsplitter tones turn on a natively ``$\chi$-matched'' interaction, as visible under qubit transition spectroscopy in numerical simulations. The $e$-$f$ transition does not appear number-split, while the $g$-$f$ transition does.
  \textbf{c}, Example initial and final Wigner functions for the calibrated T gate.
  \textbf{d}, T gate infidelity as a function of ancilla decay rate. The dynamic $\chi$-matched pulse is resistant to ancilla bit-flip induced dephasing errors events, and shows an improvement in logical gate fidelity over an equivalent fixed-$\chi$ setup. The ancilla decay rate also induces a corresponding oscillator decay through the Purcell effect, and both schemes show improvement when an ideal QEC recovery operation (see text) is applied to correct first-order oscillator decay errors.
}\label[fig]{fig:FigChiMatch}
\end{figure*}

Even with the ability to dynamically tune through the dispersive interaction space, it is difficult to combat certain deleterious effects, like the propagation of ancilla decay errors.
To fully utilize the parametric nature of the coupling and address ancilla decay, we consider a second demonstration with ``multi-tone" hybridization.
% \DKW{Need to put into the topic sentence here why one would want/need multi-tone}
If $g_\text{BS}(t)$ contains at least two frequency components, the resulting physics cannot be mapped to any effective coupling at a fixed $\Delta_\text{BS}$.
In particular, by choosing frequency components that are sufficiently far apart to prevent interference effects, one can combine different nonlinear interaction strengths from the rich landscape in~\cref{fig:Fig2}.
This allows new interaction Hamiltonians with physics beyond any non-parametric architecture.
Specifically, for a bi-chromatic parametric coupling drive with effective strengths and detunings ($g_1, \Delta_1$) and ($g_2, \Delta_2$), the effective nonlinear-interaction Hamiltonian is given by:
\begin{align}
\label{eq:nonlinearities_add}
    &H_\text{int}(g_1, \Delta_1, g_2, \Delta_2) = H_\text{int}(g_1, \Delta_1) + H_\text{int}(g_2, \Delta_2) \\
    + 
    &H_\text{int}\left(\sqrt{g_1g_2}, \frac{\Delta_1 + \Delta_2}{2}\right)\times\mathcal{O}\left[f_{\rm{periodic}}(t)
    e^{-(|\delta\Delta|T_{\text{ramp}})^2}\right]
    \nonumber.
\end{align}
The last term originates from interference effects between the two tones and can be ignored for sufficiently detuned frequency components. When interference effects are not fast rotating, the effective Hamiltonian features periodic behavior at the beat frequency of $\delta\Delta = \Delta_1 - \Delta_2$. giving rise to interesting effects like oscillating dispersive shifts and Purcell effects, which we discuss in~\cref{app:multi-tone_intereference}.

To demonstrate the utility of multi-tone operation, we use two tones that effectively cancel the dispersive shift between the cavity states and the transmon's $e-f$ transition, while preserving the shift to its $g-e$ transition.
This condition, often referred to as $\chi$-matching~\cite{PhilSergeFTDetection}, allows the use of a three-level ancilla to gain first-order tolerance to both ancilla decay and dephasing error in bosonic control pulses.
$\chi$-matching in static-coupled systems generally requires the use of a drive on the ancilla~\cite{PhilSergeFTDetection,Akshay_chi_tuning}, which can only match $\chi$ in a narrow range of Fock states, or a fine-tuned qubit frequency condition in hardware.
Here, the drives can be applied to a purpose-built coupler (see \cref{app:coupler}), apply to any qubit-resonator detuning, and the $\chi$ can in principle be matched across a large number of Fock states.
We specifically show an implementation where this two-tone coupling is used to perform a $\chi$-matched SNAP (selective Number Arbitrary Phase) gate~\cite{SNAP_2015} on the resonator, applying a logical T gate on the kitten code (see \cref{fig:FigChiMatch}a, b, c).
This logical gate is then first-order insensitive to cavity dephasing induced by ancilla decay, and the infidelity of the resonator state shows significant improvement even over a static-coupled three-level system (\cref{fig:FigChiMatch}d).
Note that the resonator inherits a Purcell decay that exists for both the static and dynamic coupling, and thus we can also compare the performance of the T gate after a round of `ideal' quantum error correction (QEC) on the resonator.
We numerically implement this QEC recovery channel by applying the operators $K_0 = \left(1 + \frac{\kappa_{\text{c, eff}}}{2} \hat{n}\right)$, unshearing the no-jump evolution, and $K_1 = \frac{1}{\sqrt{2}}(|0\rangle+|4\rangle)\langle 3| + |2\rangle\langle1|$, reversing first-order jumps.
Both static and dynamic couplings show improvement under such QEC, but even the error-corrected static-coupled system does not perform better than the raw dynamic coupling fidelity.

\begin{figure*}
    \centering
    \includegraphics[width=0.98 \textwidth]{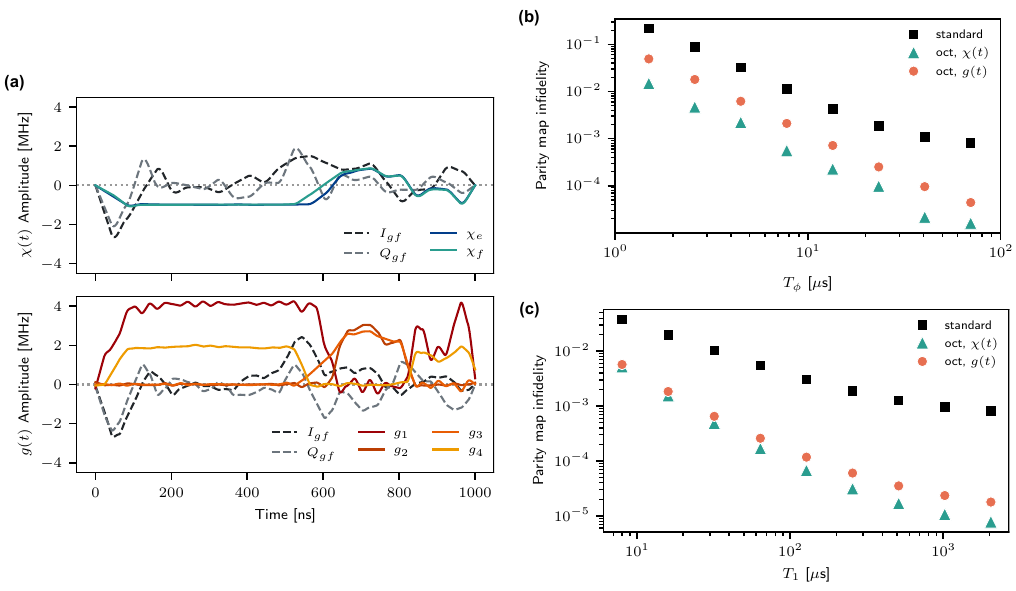}
    \caption{\label[fig]{fig:oct}\textbf{Numerically optimized parity map on the kitten code}. (a) Parity map pulse waveform. Top: Time-dependent dispersive shifts $\chi_{e}(t), \chi_{f}(t)$ and qubit drives $I_{gf}(t),Q_{gf(t)}$. Bottom: Corresponding $g_{i}(t)$ drives with $\Delta_{1}/2\pi = -25$ MHz, $\Delta_{2}/2\pi = 75$ MHz, $\Delta_{3}/2\pi = 175$ MHz, and $\Delta_{4}/2\pi = 225$ MHz, and qubit drives.  (b),(c) The dynamic pulses perform significantly better than a standard fixed-coupling approach, and are resistant to both ancilla decay ($T_{1}$) and low-frequency dephasing ($T_{\phi}$). The pure $\chi(t)$-optimized pulse represents a lower bound on infidelity that a true $g(t)$-optimized pulse can reach.}
\end{figure*}

\subsection*{Numerically optimized operations}

We have so far described 
novel control schemes enabled by
dynamic hybridization, 
including pulses that modulate the dispersive shift between positive and negative values, and enforce effective symmetries in the dispersive shifts to various ancilla states.
However, such analytically obtained pulses represent only a small fraction of the design space available to us.
Indeed, such pulses involve ramping to an effective dressed frame, with ancilla rotations enacted only once that frame has been reached.
In general, both the dispersive shifts on the oscillator and the drives on the ancilla can be modulated simultaneously. 
In this section, we explore numerically optimized pulses that can fully explore this design space, leveraging this additional freedom to improve upon piecewise analytic protocols.
In particular, we focus on improving the speed and fidelity of bosonic control pulses while retaining robust protection from ancilla errors. 
Importantly, the effects of such ancilla errors have limited many prior demonstrations of bosonic control, including the ability to perform error correction significantly beyond break-even \cite{sivak_real-time_2023}.

Here, we demonstrate this paradigm on the example of a parity measurement.
Previous work has shown that this operation can be protected from ancilla decay by using the $\ket{g},\ket{f}$ states of the ancilla and driving off-resonantly to ``$\chi$-match"; in other words, forcing $\chi_{e}=\chi_{f}$ \cite{PhilSergeFTDetection}.
In \cref{sec:echoed} we demonstrated that by echoing $\chi$ and performing a $\pi$-pulse on the ancilla, low-frequency dephasing can be mitigated.
As a showcase of the broader control space accessible through numerical optimization, we use optimal control theory to find a parity map pulse that simultaneously protects against both ancilla decay and low-frequency dephasing.
These two protections can also be combined in a piecewise analytic $g$-$f$ echoed parity map, as detailed in \cref{app:analytic_gf_echo};  however, this construction takes twice as long as a conventional parity map and twice as long as the optimized pulse considered below.

We allow the optimizer to simultaneously activate dynamic $\chi$ as well as ancilla drives, towards obtaining a $T_{1}$- and $T_{\phi}$-protected parity measurement in the case of the kitten code \cite{Michael2016}. 
We split this optimization into two steps. In the first step, we optimize the control pulses in the Hamiltonian 
\begin{align}
\label{eq:H_oct_1}
H_{\rm OCT,1}/\hbar &= \chi_{e}(t)\hat a^{\dagger}\hat a 
\ket{e}\bra{e} + \chi_{f}(t)\hat a^{\dagger}\hat a \ket{f}\bra{f} \\ \nonumber &\quad + I_{\rm gf}(t) (\ket{g}\bra{f} + \ket{f}\bra{g}) \\ \nonumber &\quad+iQ_{\rm gf}(t) (\ket{g}\bra{f} - \ket{f}\bra{g}).
\end{align}
At this stage, we optimize for obtaining the required phase evolution (a parity map) as well as for $T_{1}$ and $T_{\phi}$ protection as described in detail in \cref{app:oct}. One pulse resulting from this optimization is shown in \cref{fig:oct}(a). Notice that the optimizer chose to ``$\chi$-match" $\chi_{e}(t)=\chi_{f}(t)$ \footnote{We utilized a $\chi$-matched seed pulse, see \cref{app:oct} for details.}, helping to achieve $T_{1}$ protection. Meanwhile, the strong and always-on ancilla drives effectively perform spin locking \cite{FeiYan2013}, helping to explain the protection against ancilla dephasing. 

In the second step, $\chi_{e}(t)$ and $\chi_{f}(t)$ are together translated into four corresponding $g_{j}(t), \Delta_{j}$ drives, see \cref{app:oct} for details. That is, the pulses obtained in this first step are then used as a seed for optimizing controls in the Hamiltonian
\begin{align}
\label{eq:H_oct_2}
H_{\rm OCT,2}/\hbar &= -\frac{K}{2}\hat{q}^{\dagger}\hat{q}^{\dagger}\hat q\hat q + \frac{K}{2}\hat{q}^{\dagger}\hat{q} \\ \nonumber & \quad+ \sum_{j=1}^{4} g_{j}(t)\left(e^{-i[\Delta_{j}-K/2]t} \hat{a}^{\dagger} \hat{q} + {\rm H.~
c} \right) \\ \nonumber 
&\quad +\frac{I_{gf}(t)}{\sqrt{2}} \left( \hat{q}^2 + {\rm H.~c.} \right)
+\frac{iQ_{gf}(t)}{\sqrt{2}} \left( \hat{q}^2 - {\rm H.~c.} \right),
\end{align}
where we take the ancilla to be a Kerr oscillator, and have gone into a frame such that the $|g\rangle, |f\rangle$ levels are on resonance.
To simplify this analysis, we keep $\Delta_{j}$ static for each individual beamsplitter drive. The optimization at this stage ensures that any non-idealities obtained from translating the $\chi(t)$ drives into true detuned beamsplitter drives are mitigated, and the resulting pulse is shown in Fig.~\ref{fig:oct}(a). 
Notice that for negative $\chi_{e}(t), \chi_{f}(t)$, the drives $g_{1}(t)$ and $g_{4}(t)$ are nonzero (at $-25$ MHz and 225 MHz, respectively) while for positive $\chi_{e}(t), \chi_{f}(t)$, the drives $g_{2}(t)$ and $g_{3}(t)$ are nonzero (at $75$ MHz and 175 MHz, respectively). 

We have found empirically that splitting up the optimization in this way yields better results than simply optimizing directly on Eq.~\eqref{eq:H_oct_2}. This is due to a number of factors: (i) $g(t)$ drives tend to try to cause population swapping, while we want these drives to only weakly hybridize the modes and induce phase accumulation; (ii) in Eq.~\eqref{eq:H_oct_2} states acquire phase not only from the desired hybridization but also from the dynamic Lamb shift (see  \cref{eq:dispersive_H_w_NL}); (iii) there are an increased number of control points to optimize, slowing down the simulations. We provide the framework used for these optimizations as the open-source library \texttt{qontrol} \cite{qontrol}, which performs optimal control using \texttt{dynamiqs} \cite{dynamiqs}.

The protection for the ``$g(t)$" pulse against decay and low-frequency dephasing as compared to the standard parity-map operation is shown in \cref{fig:oct}(b) and \cref{fig:oct}(c), respectively. For the standard pulse, we employ a beamsplitter drive with 100 ns cosine ramps and take $\Delta/2\pi=-25$ MHz. 
It is important to note that when calculating the infidelity in the presence of ancilla $T_{1}$, a complication arises due to the reverse-Purcell effect. That is, upon activating a beamsplitter between the cavity and ancilla, the cavity inherits a portion of the ancilla $T_{1}$. Here, we are interested only in obtaining operations that are robust to ancilla decay: we do not expect to be able to protect against cavity decay occurring during the parity-map operation. Indeed, such a decay would likely be caught in the next measurement round. As such, when calculating the infidelity in Fig.~\ref{fig:oct} we project onto the parity subspace that the cavity state began in (importantly, this projection does not lose information about logical errors, which occur within a parity subspace).

\section{Conclusions and Discussion}
We have introduced a regime of bosonic control in which the hybridization between an oscillator and a nonlinear ancilla is itself a dynamically tunable resource.
Parametric drives control both the coupling strength and effective detuning, providing access to the multi-pole structure of the dispersive interaction and allowing its magnitude and sign to be tailored during an operation without traversing any poles directly.
To better understand the unique nature of this operating regime, we have developed concrete analytic predictions for the behavior of the effective Hamiltonian in this regime, and leverage its predictive power in our simulations.

To illustrate the utility of this control technique, we have numerically demonstrated how this freedom enables dynamical decoupling during parity measurements and ancilla-decay-tolerant logical gates with substantial improvement over conventional parity maps.
Further, we demonstrated that this control scheme is readily compatible with pulse optimization techniques common in the field.

While the exploration in this paper is purely theoretical, these protocols can be readily implemented experimentally using a parametric coupler with low nonlinearity~\cite{ChapmanSNAILBeamsplitterPaper,MaitiLINCTheory2025}, which should be used to assess their performance in realistic syndrome measurements and logical operations.
More broadly, the examples presented here explore only part of the available control space. 
Dynamic hybridization offers opportunities to develop protected gates for other bosonic encodings, tailor unwanted nonlinearities alongside the desired interaction, and exploit ancillae with different nonlinear spectra. 
Together, being able to decouple while idling and engineer protected interactions when performing gates, offers a route toward bosonic processors no longer limited by the ancillae used to control them.

\section{Acknowledgements}
We are grateful to Luigi Frunzio for helpful feedback on the manuscript. 

This research was sponsored by the Army Research Office (ARO) under grant no. W911NF-23-1-0051, and by the U.S. Department of Energy (DoE), Office of Science, National Quantum Information Science Research Centers, and Co-design for Quantum Advantage (C2QA) under contract number DE-SC0012704. Y.L. acknowledges support from the U.S. Department of Energy, Office of Science, National Quantum Information Science Research Centers, Superconducting Quantum Materials and Systems Center (SQMS), under Contract No. 89243024CSC000002. Fermilab is operated by Fermi Forward Discovery Group, LLC under Contract No. 89243024CSC000002 with the U.S. Department of Energy, Office of Science, Office of High Energy Physics.
The views and conclusions contained in this document are those of the authors and should not be interpreted as representing the official policies, either expressed or implied, of the ARO, DoE, or the U.S. Government. The U.S. Government is authorized to reproduce and distribute reprints for Governmental purposes notwithstanding any copyright notation herein. 

S.M.G. and R.J.S. are consultants and shareholders
of D-Wave Quantum.
\\
\indent A.M. conceived the central idea.
J.W.O.G. derived the detailed interaction Hamiltonian in the presence of the drive.
T.T. and J.W.O.G. derived and simulated non-adiabatic effects due to finite pulse ramps. 
A.M. and J.W.O.G. simulated pulse schemes for echoed and chi-matched bosonic control pulses, and developed the low-frequency dephasing simulation techniques. 
D.K.W. provided theoretical support and simulated numerically optimized control pulses. 
Y.L. analyzed effects from utilizing realistic couplers for this scheme. 
R.J.S. and S.M.G. provided overall guidance. All authors wrote and gave feedback on the manuscript.

% \vspace{0.25in}
% \centering
\begin{center}
\rule[0.5ex+0.25pt]{0.06\textwidth}{0.5pt}% A
\rule[0.5ex+0.pt]{0.06\textwidth}{1.pt}% B
\rule[0.5ex-0.25pt]{0.06\textwidth}{1.5pt}% C
\rule[0.5ex-0.5pt]{0.06\textwidth}{2.pt}% D
\rule[0.5ex-0.25pt]{0.06\textwidth}{1.5pt}% C 
\rule[0.5ex]{0.06\textwidth}{1.0pt}% B
\rule[0.5ex+0.25pt]{0.06\textwidth}{0.5pt}% A
\end{center}

% \vspace{0.2in}

% \clearpage
\appendix
\section{Simulations with a Coupler}
\label[app]{app:coupler}

In the main text, we consider a Hamiltonian model where the nonlinear ancilla is coupled to the linear oscillator via a programmable beamsplitter interaction, allowing for the generation of a dynamic dispersive coupling between the oscillator and the ancilla. In reality, such an interaction is enabled by a physical parametric coupler, such as the LINC~\cite{MaitiLINCTheory2025}. In this section, we model the full driven oscillator--LINC--ancilla system and verify that it realizes the interactions assumed in the main text. We calculate the parametrically activated beamsplitter strength, the dynamic dispersive shift and drive-induced frequency shifts, and the inherited oscillator decay arising from both ancilla and LINC relaxation.

% Throughout this section, we set $\hbar=1$. 
% hbars added by John on Nov 1st
The oscillator and nonlinear ancilla are described by annihilation operators $\hat a$ and $\hat q$, respectively, while the LINC mode is described by $\hat c$. Before expressing the LINC in its oscillator basis, the driven circuit Hamiltonian may be written as
\begin{align}
\hat H(t)/\hbar
={}&
\omega_a \hat a^\dagger \hat a
+\omega_q \hat q^\dagger \hat q
-\frac{K}{2}
\hat q^\dagger\hat q^\dagger\hat q\hat q
\nonumber\\
&+
i g_a'
\left(\hat a^\dagger-\hat a\right)\hat n_c
+
i g_q'
\left(\hat q^\dagger-\hat q\right)\hat n_c
\nonumber\\
&+
4\frac{E_{C c}}{\hbar}\hat n_c^2
+\frac{E_{L c}}{2\hbar}\hat\theta^2
-2\frac{E_{J c}}{\hbar}\cos\left[\frac{\varphi(t)}{2}\right]
\cos\hat\theta ,
\label{eq:full_linc_charge_hamiltonian}
\end{align}
where $\hat n_c$ and $\hat\theta$ are the conjugate charge and phase operators of the LINC. The external reduced flux is
\begin{align*}
\varphi(t)
&=
2\pi\frac{\Phi_{\mathrm{ext}}(t)}{\Phi_0}
=
\pi+\varphi_{\mathrm{AC}}(t),
\nonumber\\
\varphi_{\mathrm{AC}}(t)
&=
2\pi\frac{\Phi_{\mathrm{AC}}}{\Phi_0}
\cos(\omega_d t),
\label{eq:linc_flux_drive}
\end{align*}
where we operate at the DC flux bias $\Phi_{\mathrm{ext}}^{\mathrm{DC}}=\Phi_0/2$. At this operating point, the static Josephson contribution vanishes and the LINC is well approximated as a linear oscillator. Writing
\begin{equation}
\hat\theta
=
\theta_{\mathrm{zpf}}
\left(\hat c^\dagger+\hat c\right),
\end{equation}
we obtain
\begin{align}
\hat H(t)/\hbar
={}&
\omega_a \hat a^\dagger \hat a
+\omega_q \hat q^\dagger \hat q
-\frac{K}{2}
\hat q^\dagger\hat q^\dagger\hat q\hat q
+\omega_c\hat c^\dagger\hat c
\nonumber\\
&-
g_a
\left(\hat a^\dagger-\hat a\right)
\left(\hat c^\dagger-\hat c\right)
-
g_q
\left(\hat q^\dagger-\hat q\right)
\left(\hat c^\dagger-\hat c\right)
\nonumber\\
&+
2\frac{E_{J c}}{\hbar}
\sin\left[\frac{\varphi_{\mathrm{AC}}(t)}{2}\right]
\cos\left[
\theta_{\mathrm{zpf}}
\left(\hat c^\dagger+\hat c\right)
\right].
\label{eq:full_linc_hamiltonian}
\end{align}
When the flux is modulated near the static oscillator--ancilla detuning, the driven LINC mediates a frequency-converting interaction between the oscillator and the ancilla. Tracing out the LINC then produces the beamsplitter interaction used in Eq.~\eqref{eq:interaction_Hamiltonian}.

For the numerical calculations below, we use
\begin{align}
\omega_a/2\pi &= 4.890~\mathrm{GHz},
&
\omega_q/2\pi &= 7.808~\mathrm{GHz},
\nonumber\\
\omega_c/2\pi &= 6.500~\mathrm{GHz},
&
K/2\pi &= 100~\mathrm{MHz},
\nonumber\\
g_a/2\pi &= 120~\mathrm{MHz},
&
g_q/2\pi &= 100~\mathrm{MHz},
\nonumber\\
E_{Jc}/\hbar &= 15.84~\mathrm{GHz},
&
\theta_{\mathrm{zpf}} &= 0.248.
\label{eq:linc_parameters}
\end{align}
For the full oscillator--LINC--ancilla Floquet calculations, we retain six levels for each of the three modes. Floquet modes are assigned by their time-averaged overlap with the corresponding bare states and are tracked continuously as the drive frequency or amplitude is varied.

\subsection{Parametrically activated beamsplitter interaction}

We first calculate the effective beamsplitter strength generated by the driven LINC. For this calculation, the oscillator and ancilla are not explicitly retained in the Floquet Hilbert space. Instead, we diagonalize the periodically driven LINC and treat its couplings to the oscillator and ancilla perturbatively, following the Floquet linear-response formalism of Ref.~\cite{Yaxing_PRA_BilinearModeCoupling} [see Eq.~(30) therein]. The effective frequency-converting interaction is obtained from the first-sideband off-diagonal susceptibility of the driven LINC,
\begin{equation}
\hat H_{\mathrm{BS}}/\hbar
=
g_a g_q
\chi_c^{(1)}(\omega_a,\omega_q)
\hat a^\dagger\hat q
+
\mathrm{H.c.},
\label{eq:linc_floquet_susceptibility}
\end{equation}
where $\chi_c^{(1)}$ describes the response at the frequency shifted by one modulation quantum. We absorb the phase of this susceptibility into the phase of the flux modulation and define
\begin{equation}
g_{\mathrm{BS}}
=
g_a g_q
\left|
\chi_c^{(1)}(\omega_a,\omega_q)
\right|.
\label{eq:gbs_from_susceptibility}
\end{equation}

\begin{figure}[t]
    \centering
    \includegraphics[width=\linewidth]{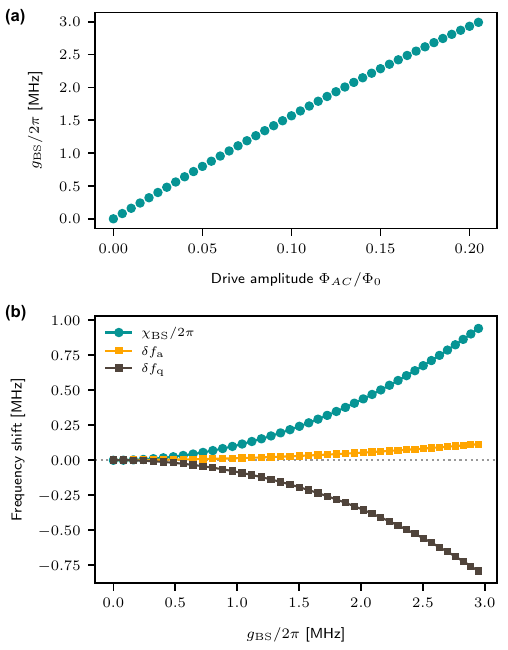}
    \caption{
    \textbf{Parametrically activated beamsplitter and dispersive strength.}
    \textbf{(a),} Effective oscillator--ancilla beamsplitter strength
    $g_{\mathrm{BS}}/2\pi$, calculated from the Floquet
    susceptibility of the driven LINC, as a function of the AC
    flux amplitude $\Phi_{\mathrm{AC}}/\Phi_0$. The oscillator and
    ancilla are treated perturbatively and are not explicitly
    included in the Floquet diagonalization used for this panel.
    \textbf{(b),} Corresponding dynamic dispersive shift $\chi_{\mathrm{BS}}/2\pi$ and
    frequency shifts $\delta f_a$ and $\delta f_q$ as a function
    of coupling $g_{\mathrm{BS}}/2\pi$, evaluated at
    $\Delta_{\mathrm{BS}}/2\pi=50~\mathrm{MHz}$. The dispersive
    shift and frequency shifts grow approximately
    quadratically with the activated beamsplitter strength.
    }
    \label[fig]{fig:linc_gbs_flux}
\end{figure}

For each applied AC flux amplitude, we numerically evaluate
$\chi_c^{(1)}$ from the driven LINC Floquet modes and use
Eq.~\eqref{eq:gbs_from_susceptibility} to obtain
$g_{\mathrm{BS}}$, as shown in \cref{fig:linc_gbs_flux}a.
The interaction is approximately linear in $\Phi_{\mathrm{AC}}$
over the range considered and reaches
$g_{\mathrm{BS}}/2\pi\simeq3~\mathrm{MHz}$ at the moderate flux
amplitude $\Phi_{\mathrm{AC}}/\Phi_0\simeq0.2$. This verifies that
the physical LINC can access the coupling strengths considered
throughout the main text.

\subsection{Dynamic dispersive shift and drive-induced frequency shifts}

We next retain all three modes and directly calculate the Floquet spectrum of Eq.~\eqref{eq:full_linc_hamiltonian}. We denote by $\epsilon_{\ell,m,n}$ the unwrapped quasienergy of the Floquet mode continuously connected to the bare state
\begin{equation}
\ket{\ell,m,n}
=
\ket{\ell}_c\ket{m}_a\ket{n}_q,
\end{equation}
where the indices label the LINC, oscillator, and ancilla occupations, respectively. The drive-induced shifts of the oscillator and ancilla frequencies are defined as
\begin{align}
\delta f_a
&=
\frac{
\epsilon_{0,1,0}-\epsilon_{0,0,0}
}{2\pi}
-\frac{\bar\omega_a}{2\pi},
\label{eq:linc_cavity_shift}
\\
\delta f_q
&=
\frac{
\epsilon_{0,0,1}-\epsilon_{0,0,0}
}{2\pi}
-\frac{\bar\omega_q}{2\pi},
\label{eq:linc_ancilla_shift}
\end{align}
where $\bar\omega_a$ and $\bar\omega_q$ are the dressed oscillator and ancilla frequencies without drive. The dynamic dispersive shift is extracted from the quasienergy combination
\begin{equation}
\frac{\chi_{\mathrm{BS}}}{2\pi}
=
\frac{
\epsilon_{0,1,1}
-\epsilon_{0,1,0}
-\epsilon_{0,0,1}
+\epsilon_{0,0,0}
}{2\pi}.
\label{eq:linc_floquet_chi}
\end{equation}

\begin{figure}[ht]
    \centering
    \includegraphics[width=\linewidth]{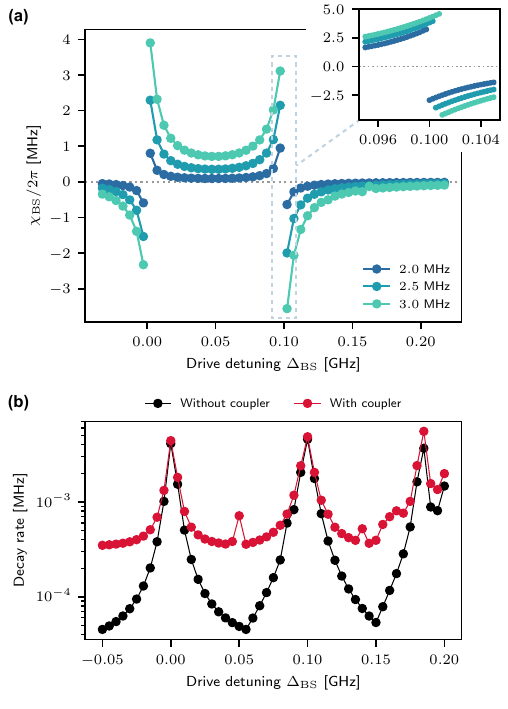}
    \caption{
    \textbf{Dynamic dispersive shift and Purcell limits obtained from the full
    oscillator--LINC--ancilla Floquet spectrum.}
    \textbf{(a),} Oscillator--ancilla dispersive shift
    $\chi_{\mathrm{BS}}/2\pi$ as a function of the bare drive
    detuning $\Delta_{\mathrm{BS}}/2\pi$ for flux amplitudes
    corresponding to $g_{\mathrm{BS}}/2\pi=2$, $2.5$, and
    $3~\mathrm{MHz}$. Inset, enlarged view of the pole near
    $\Delta_{\mathrm{BS}}/2\pi=100~\mathrm{MHz}$. The pole moves
    with increasing drive amplitude because of the drive-induced
    oscillator and ancilla frequency shifts, as visible in the inset.
    \textbf{(b),} Pointwise worst-case decay rate
    $\kappa_{1,\mathrm{max}}$ as a function of the bare
    beamsplitter detuning. Black points show the result with ancilla
    relaxation alone, while red points include both ancilla and
    LINC relaxation. For each curve, the maximum is taken over the
    ancilla initially occupying $\ket{g}$, $\ket{e}$, or $\ket{f}$.
    The broad resonances near the drive-shifted values of
    $\Delta_{\mathrm{BS}}=0$, $K$, and $2K$ are the
    state-conditioned single-photon Purcell resonances associated
    with these three initial ancilla states. LINC relaxation adds an
    off-resonant decay floor and produces its largest correction
    near the higher-lying resonance. For the LINC relaxation
    channel, $T_{1,c}=20~\mu\mathrm{s}$.
    }
    \label[fig]{fig:linc_chi_detuning}
\end{figure}

We define the bare beamsplitter detuning, as in the main text, by
\begin{equation}
\Delta_{\mathrm{BS}}
=
\bar\omega_q-\bar\omega_a-\omega_d.
\label{eq:linc_bare_detuning}
\end{equation}
\cref{fig:linc_chi_detuning}a shows the dispersive shift as a function of this detuning for flux amplitudes corresponding to $g_{\mathrm{BS}}/2\pi=2$, $2.5$, and $3~\mathrm{MHz}$. The full three-mode calculation reproduces the multi-pole dispersive landscape obtained from the effective Hamiltonian in the main text. In particular, $\chi_{\mathrm{BS}}$ changes sign when the drive crosses the oscillator--ancilla resonances.

The pole positions are shifted from the bare conditions $\Delta_{\mathrm{BS}}=0$ and $\Delta_{\mathrm{BS}}=K$ because the modulation also changes the oscillator and ancilla frequencies. To leading order, and neglecting the smaller drive-induced correction to $K$, the pole near the ancilla $e$--$f$ transition satisfies
\begin{equation}
\frac{\Delta_{\mathrm{BS}}^{(ef)}}{2\pi}
\simeq
\frac{K}{2\pi}
+\delta f_a-\delta f_q.
\label{eq:linc_shifted_ef_pole}
\end{equation}
The zoomed panel of \cref{fig:linc_chi_detuning}a shows this displacement explicitly. Increasing the flux amplitude moves the pole progressively away from the bare value $K/2\pi=100~\mathrm{MHz}$. Thus, the laboratory drive frequency corresponding to a specified effective detuning must include the drive-induced frequency shifts of the full circuit.

\cref{fig:linc_gbs_flux}b shows
$\chi_{\mathrm{BS}}$, $\delta f_a$, and $\delta f_q$ as the
beamsplitter strength is increased at the fixed bare detuning
$\Delta_{\mathrm{BS}}/2\pi=50~\mathrm{MHz}$. All three quantities
vanish in the absence of the flux modulation and grow approximately
quadratically with $g_{\mathrm{BS}}$ in the perturbative regime. The
ancilla shift is negative while the oscillator shift is positive for
the parameters considered here. Their difference reduces the effective oscillator–ancilla detuning at a fixed drive frequency, shifting the resonance positions toward larger values of the bare detuning plotted in \cref{fig:linc_chi_detuning}a. Importantly, the frequency shifts
remain small compared with the static mode frequencies and can be
calibrated through the room-temperature drive frequency.

\subsection{Inherited oscillator decay}

We finally use the Floquet--Markov formalism to quantify the
single-photon decay inherited by the oscillator from relaxation of
the ancilla and the LINC. These channels are represented
by the jump operators
\begin{equation}
\hat L_q=\sqrt{\gamma_q}\,\hat q,
\qquad
\hat L_c=\sqrt{\gamma_c}\,\hat c,
\label{eq:linc_collapse_operators}
\end{equation}
where $\gamma_q=1/T_{1,q}$ and $\gamma_c=1/T_{1,c}$. We use the
label $r\in\{q,c\}$ to distinguish the ancilla- and LINC-relaxation
channels.

Within the Floquet--Markov formalism, the transition rate from
Floquet mode $\alpha$ to Floquet mode $\beta$ induced by relaxation
channel $r$ is
\begin{equation}
\Gamma_{\alpha\rightarrow\beta}^{(r)}
=
\sum_k
\left|
L_{\beta\alpha}^{(r,k)}
\right|^2,
\label{eq:floquet_markov_rate}
\end{equation}
where
\begin{equation}
L_{\beta\alpha}^{(r,k)}
=
\frac{1}{T}
\int_0^T dt\,
e^{-ik\omega_d t}
\mel{u_\beta(t)}{\hat L_r}{u_\alpha(t)}
\label{eq:floquet_matrix_element}
\end{equation}
is the $k$-th Fourier component of the jump-operator matrix element
between Floquet modes. The sum over $k$ accounts for transitions
involving the exchange of integer numbers of drive photons. In the simulations, we take
$T_{1,c}=1/\gamma_c=20~\mu\mathrm{s}$, and 
$T_{1,q}=1/\gamma_q=100~\mu\mathrm{s}$.

For each oscillator occupation $m$ and ancilla occupation $n$, we
define the total outgoing rate from the Floquet mode
$\alpha_{m,n}$ connected to the bare state $\ket{0,m,n}$ as
\begin{equation}
\Gamma_{m,n}^{(r)}
=
\sum_{\beta\neq\alpha_{m,n}}
\Gamma_{\alpha_{m,n}\rightarrow\beta}^{(r)}.
\label{eq:total_floquet_escape}
\end{equation}
Here, the first index of $\ket{0,m,n}$ denotes the LINC occupation,
while the second and third indices denote the oscillator and ancilla
occupations. The rate $\Gamma_{0,n}^{(r)}$ contains relaxation of an
ancilla that is initially excited, together with drive-assisted
transitions that occur without an oscillator photon. We therefore
define the additional escape rate associated with one oscillator
photon as
\begin{equation}
\kappa_{1,n}^{(r)}
=
\Gamma_{1,n}^{(r)}
-
\Gamma_{0,n}^{(r)}.
\label{eq:kappa_one_photon}
\end{equation}
This differential escape rate is the Floquet--Markov analogue of
the state-dependent no-jump decay rate used in Fig.~2(b) of the
main text.

We consider the ancilla initially in the $\ket{g}$, $\ket{e}$, or
$\ket{f}$ state, corresponding to $n=0$, $1$, or $2$, and define the
pointwise worst-case inherited decay rate
\begin{equation}
\kappa_{1,\mathrm{max}}^{(r)}(\omega_d)
=
\max_{n\in\{0,1,2\}}
\kappa_{1,n}^{(r)}(\omega_d).
\label{eq:kappa_one_photon_worst_case}
\end{equation}
The broad Purcell resonances near
\begin{equation}
\Delta_{\mathrm{BS}}\simeq nK
\end{equation}
arise from the resonant exchange processes
\begin{equation}
\ket{1,n}
\longleftrightarrow
\ket{0,n+1}.
\label{eq:state_conditioned_purcell}
\end{equation}
Consequently, the resonances near $\Delta_{\mathrm{BS}}=0$, $K$,
and $2K$ correspond to the ancilla initially occupying
$\ket{g}$, $\ket{e}$, and $\ket{f}$, respectively. The curve in
Fig.~2(b), and the corresponding curves below, should therefore be
understood as worst-case envelopes over these three ancilla states.
If the ancilla remains in $\ket{g}$, only the resonance near
$\Delta_{\mathrm{BS}}=0$ is present. In the full driven circuit, the
positions of these resonances are displaced from the bare values by
the drive-induced frequency shifts discussed above.

For independent ancilla and LINC relaxation baths, the transition
rates add for a fixed initial Floquet mode,
\begin{equation}
\Gamma_{m,n}^{(q+c)}
=
\Gamma_{m,n}^{(q)}
+
\Gamma_{m,n}^{(c)},
\label{eq:combined_floquet_escape}
\end{equation}
and hence
\begin{equation}
\kappa_{1,n}^{(q+c)}
=
\kappa_{1,n}^{(q)}
+
\kappa_{1,n}^{(c)}.
\label{eq:combined_kappa_one_photon}
\end{equation}
We compare the pointwise maxima
\begin{equation}
\kappa_{1,\mathrm{max}}^{(q)}
=
\max_{n\in\{0,1,2\}}
\kappa_{1,n}^{(q)}
\end{equation}
and
\begin{equation}
\kappa_{1,\mathrm{max}}^{(q+c)}
=
\max_{n\in\{0,1,2\}}
\left[
\kappa_{1,n}^{(q)}
+
\kappa_{1,n}^{(c)}
\right].
\end{equation}
The maximum in the second expression is taken only after the two
bath contributions have been combined. Because the maximizing
ancilla state can differ between the ancilla-only and combined-loss
calculations, the difference
$\kappa_{1,\mathrm{max}}^{(q+c)}
-\kappa_{1,\mathrm{max}}^{(q)}$
should not generally be interpreted as the LINC-induced decay rate
for a single fixed ancilla state.

\cref{fig:linc_chi_detuning}b compares the worst-case inherited
oscillator decay obtained from ancilla relaxation alone with that
obtained when both ancilla and LINC relaxation are included. The
large resonant enhancements near the state-conditioned
oscillator--ancilla poles are primarily generated by ancilla
relaxation. Including LINC relaxation leaves the first two principal
resonances nearly unchanged, while adding an approximately
off-resonant decay floor and a more visible correction near the
higher-lying resonance around
$\Delta_{\mathrm{BS}}/2\pi\simeq180~\mathrm{MHz}$. Away from the
resonances, the relative contribution of LINC relaxation can become
larger because the ancilla-induced Purcell rate is strongly
suppressed, although the absolute inherited decay remains much
smaller than at the resonant poles.
These results show that the principal resonant Purcell features of
the dynamic oscillator--ancilla interaction remain dominated by ancilla
relaxation.
The finite lifetime of the physical LINC provides a
smaller, but non-negligible, additional decay channel and determines
the off-resonant decay floor of the full circuit.

\section{Dispersive Hamiltonian to Fourth Order}
\label[app]{app:PertTheory}
To accurately model the intricacies of the dispersive regime in the limit of $g < \Delta \sim K$, we go beyond the well-known second order approximation of the dispersive Hamiltonian. Here, we provide a derivation of the full effective Hamiltonian of a linear oscillator coupled to a Kerr oscillator up to fourth order in the small parameter $g/\Delta$.
With this derivation we present, to our knowledge, the first explicit expressions for the dispersive shift nonlinearities $\chi_e^\prime$ and $\chi_f^\prime$ in this system. We follow closely the framework of Zhu et al.~\cite{Zhu2013Guanyu4thOrder}, where the authors provide a detailed prescription for deriving effective Hamiltonians up to high order, without needing to evaluate nested commutators. One instead computes energy corrections using standard Rayleigh-Schr\"odinger perturbation theory, then infers how those corrections would have arisen from a diagonal effective Hamiltonian.
% however we crucially remedy what we believe to be a typo in the first half of equation 19. We do not question the integrity of any results in ref.~\cite{Zhu2013Guanyu4thOrder} beyond this typesetting error. 
For completeness, we work through the derivation explicitly and highlight a typo in Ref.~\cite{Zhu2013Guanyu4thOrder}, fixed here.

To model \cref{eq:detuned_bs_w_NL} in the perturbation theory framework of Ref. \cite{Zhu2013Guanyu4thOrder} 
% \DKW{Note that this is not actually what Eq.(3) looks like. Does it make sense to rewrite actually using eq(3)? this would explicitly use delta, and the spectrum won't depend on $n$},
we write the Hamiltonian as $H=H_{0}+V$, where
% we express the diagonal part of the Hamiltonian as a harmonic oscillator and a Kerr oscillator
\begin{equation}
    H_0/\hbar = \Delta \hat{q}^\dagger \hat{q} - \frac{K}{2}q^\dagger q^\dagger q q,
\end{equation}
is the diagonal bare Hamiltonian of a harmonic oscillator coupled to a Kerr oscillator, and
\begin{equation}
    V/\hbar = g (a^\dagger q + a q^\dagger),
\end{equation}
is the term we treat perturbatively. From $H_{0}$, we obtain the bare energies
\begin{equation}
    \frac{E^{(0)}_{n,l}}{\hbar} = l\Delta - \frac{K}{2}l(l - 1),
\end{equation}
written in terms of excitation numbers $n,l$ of the harmonic and Kerr oscillators, respectively, where the bare energies are independent of $n$ in this frame. 
Our goal is to find the corrections to the dressed energy levels due to the perturbation $V$.

By noting that $V$ flips ancilla excitation parity while $H_0$ preserves it,
% and that the perturbation only couples nearest-neighbor energy levels in the multi-level ladder \DKW{What is this ladder? presumably you're referring to something from Guanyu's paper, but the reader doesn't see it here}, 
we see that all odd-order corrections to the energies are zero. 
The leading-order corrections occur at second order, yielding the well-known result
\begin{align}
    E_{n,l}^{(2)} &= \frac{\vert\bra{n,l}V\ket{n+1,l-1}\vert^2}{E^{(0)}_{n,l} - E^{(0)}_{n+1,l-1}}
    +\frac{\vert\bra{n,l}V\ket{n-1,l+1}\vert^2}{E^{(0)}_{n,l} - E^{(0)}_{n-1,l+1}} \\
    &= \left(\frac{g^2(l)(n+1)}{\Delta - (l-1)K} - \frac{g^2(l+1)(n)}{\Delta - lK}\right)\hbar.
\end{align}
% \DKW{Comment here about why second order isn't good enough, otherwise jump to fourth order is abrupt.}
While this conventional result models the system quite well in the typical dispersive regime ($g \sim K \ll \Delta$), it is insufficient for resolving the finer structure of the spectrum. We must extend the analysis to higher orders to capture vital corrections to the $\ket{f}$-state dispersive shift, alongside effects like the inherited oscillator Kerr $K_{osc}$ and the nonlinearity of the dispersive shift $\chi_e'$.

The fourth-order energy corrections have the general form \cite{Zhu2013Guanyu4thOrder}
\begin{equation}
    E_{n,l}^{(4)} = \sum_{\alpha, \beta, \gamma \neq (n,l)} \frac{V_{(n,l)\gamma}V_{\gamma\beta}V_{\beta\alpha}V_{\alpha(n,l)}}{\hbar^3\Delta_{\alpha}\Delta_{\beta}\Delta_{\gamma}} - E_{n,l}^{(2)} \sum_{\gamma \neq (n,l)} \frac{|V_{(n,l)\gamma}|^2}{\hbar^2\Delta_{\gamma}^2},
\end{equation}
where $\alpha, \beta$, and $\gamma$ index the intermediate joint states $\ket{n',l'}$, and the energy denominators are defined as $\hbar\Delta_{\alpha} \equiv E_{n,l}^{(0)} - E_{\alpha}^{(0)}$. 
Leveraging the single-photon exchange nature of the perturbation, we simplify this sum to
\begin{align}
\label{eq:pert_fourth_order_corrections}
    &E_{n,l}^{(4)} = 
    \Bigg[
    \frac{\left\vert\bra{n,l}V\ket{n+1,l-1}\right\vert^2
    \left\vert\bra{n+1,l-1}V\ket{n+2,l-2}\right\vert^2}
    {\left(E^{(0)}_{n,l} - E^{(0)}_{n+2,l-2}\right)\left(E^{(0)}_{n,l} - E^{(0)}_{n+1,l-1}\right)^2} \nonumber
     \\
    &+ 
    \frac{\left\vert\bra{n,l}V\ket{n-1,l+1}\right\vert^2
    \left\vert\bra{n-1,l+1}V\ket{n-2,l+2}\right\vert^2}
    {\left(E^{(0)}_{n,l} - E^{(0)}_{n-2,l+2}\right)\left(E^{(0)}_{n,l} - E^{(0)}_{n-1,l+1}\right)^2}\Bigg] \nonumber\\
    &-E_{n,l}^{(2)} \left[
    \frac{\vert\bra{n,l}V\ket{n+1,l-1}\vert^2}{\left(E^{(0)}_{n,l} - E^{(0)}_{n+1,l-1}\right)^2}
    +\frac{\vert\bra{n,l}V\ket{n-1,l+1}\vert^2}{\left(E^{(0)}_{n,l} - E^{(0)}_{n-1,l+1}\right)^2}
    \right] \\ \nonumber 
    &\quad\equiv E_{n,l}^{(4)(I)}+E_{n,l}^{(4)(II)}.
\end{align}
We see that the fourth-order correction consists of two types of terms: 'double jumps' that involve states two levels away (the first bracket of \cref{eq:pert_fourth_order_corrections}), as well as repeated 'single jumps', shown in the second bracket. 
Simplifying the individual terms, we obtain
\begin{align}
    \frac{E_{n,l}^{(4)(I)}}{\hbar} &=  
    \Big[\frac{g^4 (n+1)(n+2)(l)(l-1)}
    {(2\Delta - (2l-3)K)(\Delta - (l-1)K)^2} \nonumber \\
    &- 
    \frac{g^4 n(n-1)(l+1)(l+2) }
    {(2\Delta  - (2l+1)K )(\Delta - lK)^2}\Big], \\
    \label{eq:pert_fourth_order_fixed_part}
    \frac{E_{n,l}^{(4)(II)}}{\hbar} &=  
    -\left[
    \frac{g^2(l)(n+1)}{\Delta - (l-1)K} - \frac{g^2(l+1)(n)}{\Delta - lK}
    \right]\nonumber \\
    &\times\left[
    \frac{g^2(l)(n+1)}{\left(\Delta - (l-1)K\right)^2}
    +\frac{g^2(l+1)(n)}{\left(\Delta - lK\right)^2}
    \right].
\end{align}
Note that the last term of \cref{eq:pert_fourth_order_fixed_part} has a positive sign, differing from Eq. 19 of ref.~\cite{Zhu2013Guanyu4thOrder}.

With this error remedied and both the second- and fourth-order corrections to the energies found, 
We now compute the dressed energies up to fourth order via 
\begin{equation}
    E_{n,l} = E_{n,l}^{(0)} +
    E_{n,l}^{(2)} + 
    E_{n,l}^{(4)(I)} + E_{n,l}^{(4)(II)} + \mathcal{O}\left(\frac{g}{\Delta}\right)^6,
\end{equation}
given as a function of oscillator and transmon excitations.
To extract out the Hamiltonian that properly describes this spectrum, we compute coefficients term-by-term using the appropriate energy differences 
% \vspace{.2in}

\begin{figure*}[ht]
    \centering\includegraphics[width=2.0\columnwidth]{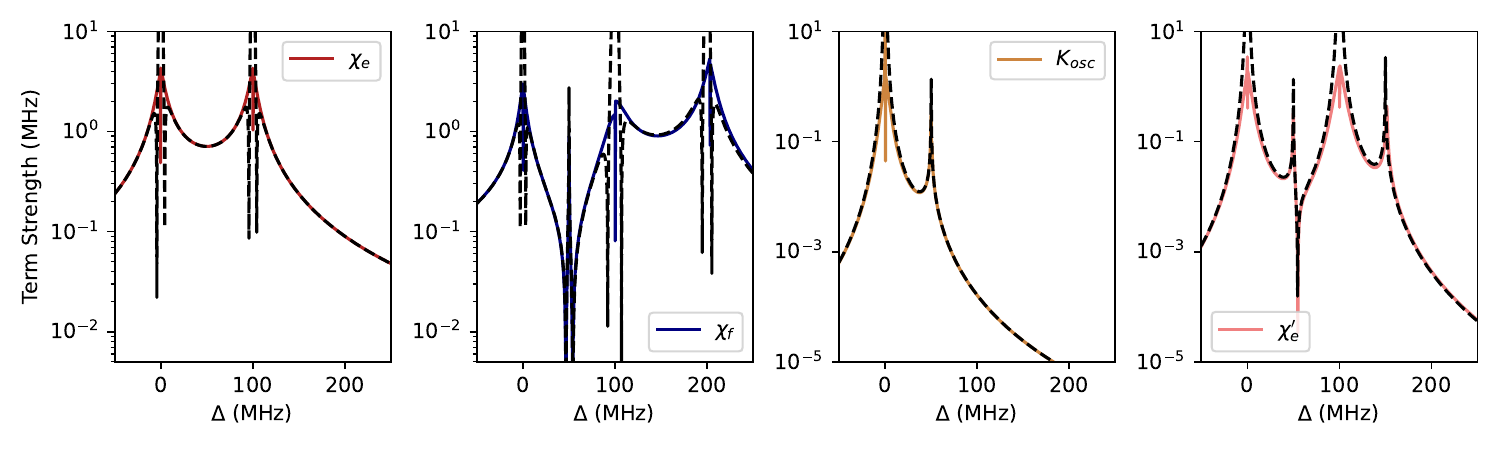}
    \caption{\label[fig]{fig:dispersive_Ham_appendix}\textbf{Comparing analytics and numerics}. Nonlinearities of the dressed system are computed via direct diagonalization of a transmon with {$E_c = 96.315$ MHz, $E_J=48.26$ GHz} (yielding a 100 MHz Kerr) and compared with the analytic expressions found in \cref{eq:full_analytic_dispersive_expressions}, using $g=3$ MHz, $K=100$ MHz. The two methods have remarkable agreement, only diverging at half-integer multiples of the ancilla Kerr nonlinearity, when the oscillator is directly on resonance with an ancilla transition and perturbative methods break down. 
    }
\end{figure*}

\begin{subequations} \label{eq:effective_H_differences}
\begin{align}
    \hbar\tilde{\omega}_{a} &= [E_{1,0} - E_{0,0}] \\
    \hbar\tilde{\omega}_{q} &= [E_{0,1} - E_{0,0}] \\
    -\hbar\tilde{K} &= [E_{0,2} - E_{0,1}] - [E_{0,1} - E_{0,0}] \\
    \hbar K_{osc} &= [E_{2,0} - E_{1,0}] - [E_{1,0} - E_{0,0}] \\
    \hbar\chi_{e} &= [E_{1,1} - E_{0,1}] - [E_{1,0} - E_{0,0}] \\
    \hbar\chi_{f} &= [E_{1,2} - E_{0,2}] - [E_{1,0} - E_{0,0}] \\
    \hbar\chi_{e}^\prime &= \big([E_{2,1} - E_{1,1}] - [E_{2,0} - E_{1,0}]\big) \\
    &- \big([E_{1,1} - E_{0,1}] - [E_{1,0} - E_{0,0}]\big) \nonumber \\
    \hbar\chi_{f}^\prime &= \big([E_{2,2} - E_{1,2}] - [E_{2,0} - E_{1,0}]\big) \\
    &- \big([E_{1,2} - E_{0,2}] - [E_{1,0} - E_{0,0}]\big).\nonumber
\end{align}
\end{subequations}
Using these energy differences, we fully capture the spectrum of $E_{n,l}$ with the Hamiltonian
\begin{align}
\label{eq:full_analytic_dispersive_Hamiltonian}
    H_{\text{eff}}/\hbar &= \tilde{\omega}_a a^\dagger a + \frac{K_{osc}}{2}a^\dagger a (a^\dagger a - 1) \\
    &+ \tilde{\omega}_q q^\dagger q - \frac{\tilde{K}}{2}q^\dagger q (q^\dagger q - 1)  \nonumber\\
    &+ [\chi_{e} + \frac{\chi_{e}^\prime}{2} (a^\dagger a - 1)]a^\dagger a \ket{e}\bra{e}  \nonumber\\
    &+ [\chi_{f} + \frac{\chi_{f}^\prime}{2} (a^\dagger a - 1)]a^\dagger a \ket{f}\bra{f} \nonumber.
\end{align}
Here, both the oscillator and ancilla experience a Lamb shift and a modification to their Kerr nonlinearity. 
Additionally, we see that the dispersive shift between them has a photon number dependence, giving rise to a nonlinear dispersive shift as a function of oscillator occupation.
The explicit functional form of these coefficients are given by

\begin{widetext} % Wide final equations begin here
\begin{subequations}
\label{eq:full_analytic_dispersive_expressions}
\eqpair{0.42}{%
    \tilde{\omega}_a = \omega_a - \frac{g^2}{\Delta} + \frac{g^4}{\Delta^3}
}{0.55}{%
    \tilde{\omega}_q = \omega_q + \frac{g^2}{\Delta} - \frac{g^4}{\Delta^3}
}

\eqpair{0.62}{%
    \tilde{K} = K - \frac{2g^2 K}{\Delta (\Delta - K)}
    - \frac{2g^4\left((K - 2\Delta)(K - \Delta)^3 - 2\Delta^4\right)}
           {\Delta^3 (K - 2\Delta)(K - \Delta)^3}
}{0.35}{%
    K_{\rm osc} = -\frac{2g^4K}{\Delta^3(2\Delta- K)}
}
% \begin{equation}
%         K_{\rm osc} = -\frac{2g^4K}{\Delta^3(2\Delta- K)}.
% \end{equation}
\begin{equation}
    \chi_{e} =         \frac{-2g^2 K }{\Delta (\Delta - K)} 
        +\frac{4g^4K(K^2 + 2\Delta(-K + \Delta))}{\Delta^3(-K + \Delta)^3}
\end{equation}
% \begin{equation}
%     \chi_{f} - \chi_{e} =             \frac{-2g_{BS}^2K(K + \Delta)}{(K - \Delta)(2K - \Delta)\Delta}
%             +2g_{BS}^4\frac{K(12K^6 - 62K^5\Delta + 135K^4\Delta^2 - 161K^3\Delta^3 + 73K^2\Delta^4 + 3K\Delta^5 - 8\Delta^6)}
%             {(K - 2\Delta)\Delta^3(2K^2 - 3K\Delta + \Delta^2)^3}
% \end{equation}
\begin{equation}
    \chi_{f} =  -\frac{2 g^{2}K\left(2\Delta-K\right)}
{\Delta\,\left(\Delta-2K\right)\,\left(\Delta-K\right)}
\;+\;
\frac{2 g^{4}\,\left(4K^{7}-26K^{6}\Delta+69K^{5}\Delta^{2}-97K^{4}\Delta^{3}+103K^{3}\Delta^{4}-63K^{2}\Delta^{5}+16K\Delta^{6}\right)}
{\Delta^{3}\,\left(2\Delta-K\right)\,\left(2K^{2}-3K\Delta+\Delta^{2}\right)^{3}}
\end{equation}
\begin{equation}
    \chi_{e}' =                \frac{4g^4K^2(-3K^3 + 11K^2\Delta - 15K\Delta^2 + 9\Delta^3)}{\Delta^3
        (K - 2\Delta)(3K - 2\Delta)(K - \Delta)^3}
\end{equation}
% Or alternatively,
% \begin{equation}
%     \chi_{ge}' =                g^4 \left( -\frac{4}{\Delta^3} 
% - \frac{4}{(K - 2 \Delta) \Delta^2} 
% + \frac{8}{(-K + \Delta)^3} 
% + \frac{12}{(3K - 2\Delta)(-K + \Delta)^2} 
% - \frac{4}{\Delta (-K + \Delta)^2} 
% + \frac{4}{\Delta^2 (-K + \Delta)} \right)
% \end{equation}
% \begin{equation}
%     \chi_{f}' - \chi_{e}' =         \frac{4g^4K^2(60K^7 - 334K^6\Delta + 799K^5\Delta^2 - 1074K^4\Delta^3 + 734K^3\Delta^4 - 182K^2\Delta^5 - 
%     33K\Delta^6 + 18\Delta^7)}
%     {\Delta^3(K - 2\Delta)(K - \Delta)^3(3K - 2\Delta)(2K - \Delta)^3(5K - 2\Delta)}
% \end{equation}
% Or alternatively,
% \begin{equation}
%     \chi_{ef}' =  g^4 \left( 
% \frac{2}{\Delta^3} 
% + \frac{18}{(-2K + \Delta)^3} 
% + \frac{24}{(5K - 2\Delta)(-2K + \Delta)^2} 
% - \frac{16}{(-K + \Delta)^3} 
% - \frac{12}{(3K - 2\Delta)(-K + \Delta)^2} 
% + \frac{4}{\Delta (-K + \Delta)^2} 
% + \frac{12}{(-2K + \Delta)(-K + \Delta)^2} 
% - \frac{4}{\Delta^2 (-K + \Delta)} 
% - \frac{12}{(-2K + \Delta)^2 (-K + \Delta)} 
% + \frac{4}{(-K + \Delta)^2 (-K + 2\Delta)} 
% \right)
% \end{equation}
\begin{equation}
    \chi_{f}' =         -\frac{4g^4K^2(20K^6 - 98K^5\Delta + 201K^4\Delta^2 - 223K^3\Delta^3 + 169K^2\Delta^4 - 81K\Delta^5 + 18\Delta^6)}
        {\Delta^3(K - 2\Delta)(K - \Delta)^3(2K - \Delta)^3(5K - 2\Delta)}.
\end{equation}
% Or alternatively,
% \begin{equation}
%     \chi_{gf}' =  g^4 \left( 
% -\frac{2}{\Delta^3} 
% - \frac{4}{(K - 2\Delta) \Delta^2} 
% + \frac{18}{(-2K + \Delta)^3} 
% + \frac{24}{(5K - 2\Delta)(-2K + \Delta)^2} 
% - \frac{8}{(-K + \Delta)^3} 
% + \frac{12}{(-2K + \Delta)(-K + \Delta)^2} 
% - \frac{12}{(-2K + \Delta)^2 (-K + \Delta)} 
% + \frac{4}{(-K + \Delta)^2 (-K + 2\Delta)} 
% \right)
% \end{equation}
\end{subequations}
\end{widetext}
While some of these expressions are quite familiar and compact, the combinatorics of higher-order perturbation theory gives rise to highly non-trivial level shifts.

To confirm the accuracy of our analytic model, we compare with a direct diagonalization of \cref{eq:detuned_bs_w_NL} as shown in \cref{fig:dispersive_Ham_appendix}. 
We find excellent agreement with numerics on a log-scale, with deviations only present near single- and two-photon resonances where the perturbative method breaks down. 
This refined understanding of the dispersive Hamiltonian allows us to precisely predict the nonlinearities activated by a dynamic parametric coupling. With concrete analytic predictions, we can more efficiently explore the full parameter space to find desirable operating points, including those where the terms of \cref{eq:full_analytic_dispersive_expressions} do not vary together, so that one can be tuned largely independently of the others.
Further, these expressions guide the search through the enlarged multi-tone parameter space, indicating in advance which combinations of detunings will produce a given set of nonlinearities.
Assuming a beamsplitter coupling, these expressions apply to any linear oscillator dispersively coupled to a Kerr oscillator, and so may be of use beyond the parametric setting considered here. Further, the explicit analytic technique shown here readily applies to more exotic nonlinear circuits \cite{Zhu2013Guanyu4thOrder}.

\section{Speed Limit Derivation}
\label[app]{app:speed_limit}
Here, we outline our approach to define the speed limit for the adiabatic ramp for the dynamically activated dispersive interaction. As established in \cref{sec:speed_limit}, the ramp must be sufficiently slow such that the inertial term is much smaller than the energy scale of the level-splitting defined by the drive, a condition captured in,
\begin{align}
|\dot{\theta}(t)|\ll\sqrt{(2g_{\rm{BS}}(t))^2 + \Delta_{\rm{BS}}(t)^2}.
\end{align}
This inequality can be transformed by introducing the definition of the mixing angle $\theta(t)=\frac{1}{2}\arctan\left(\frac{2g_{\rm{BS}}(t)}{\Delta_{\rm{BS}}(t)}\right)$,
\begin{align}
\Bigl|\frac{\dot{g}_{\rm_{BS}}(t)}{\Delta_{\rm{BS}}}\Bigr|\frac{1}{1+\bigl(\frac{2g_{\rm{BS}}(t)}{\Delta_{\rm{BS}}}\bigr)^2}\ll\vert\Delta_{\rm{BS}}\vert\sqrt{\left(1+\left\vert\frac{2g_{\rm{BS}}(t)}{\Delta_{\rm{BS}}}\right\vert^2\right)}.
\end{align}
Applying the assumption that $\it{g}_{\rm{BS}}(t) \ll \Delta_{\rm{BS}}$ simplifies this expression, leading to,
\begin{align}
\Bigl|\frac{\dot{g}_{\rm_{BS}}(t)}{\Delta_{\rm{BS}}}\Bigr|\ll\vert\Delta_{\rm{BS}}\vert,
\end{align}
To analyze this condition further, we split the left-hand side for convenience, as shown in,
\begin{align}
\Bigl|\frac{\dot{g}_{\rm_{BS}}(t)}{g}\Bigr|\Bigl|\frac{g}{\Delta_{\rm{BS}}}\Bigr|\ll\vert\Delta_{\rm{BS}}\vert,
\end{align}
where $g=\rm{max}\{\it{g}_{\rm{BS}}(t)\}$.
Finally, given that we remain in the $g \ll \Delta_{\rm{BS}}$ regime, we can conclude that $\Bigl|\frac{\dot{g}_{\rm_{BS}}(t)}{g}\Bigr|$ should be equivalent to $\Delta_{\rm{BS}}$ or less. From this observation, we defined the speed limit in \cref{fig:Fig3}c (dotted black line) by the ramp time that satisfies $\rm{max}\Bigl|\frac{\dot{\it{g}}_{\rm_{BS}}(t)}{\it{g}}\Bigr|=\Bigl|\Delta_{\rm{BS}}\Bigr|$.

\section{Ancilla-state dependence of the adiabatic condition}
\label[app]{app:ancilla-init}

In the main text, we examined how quickly a dynamic $\chi$ can be activated by evaluating the state-independent process infidelity $\varepsilon(t_{\rm{r}})$ over the truncated transmon-cavity Hilbert space [Eq.~(11) and \cref{fig:Fig3}(b)]. That two-dimensional map revealed multiple sharp resonant features, or poles, where the ramp fidelity is significantly degraded due to the multi-level nature of the transmon ancilla. Because the process infidelity averages over all initial states in the subspace, all active resonances appear simultaneously in Fig.~\ref{fig:Fig3}(b). To clarify the physical origin of this multi-pole structure and understand how the adiabatic condition depends on the ancilla state during bosonic control, we decompose the analysis by evaluating the non-adiabatic error for specific initial ancilla states.

\begin{figure*}[ht]
    \centering\includegraphics[width=2.0\columnwidth]{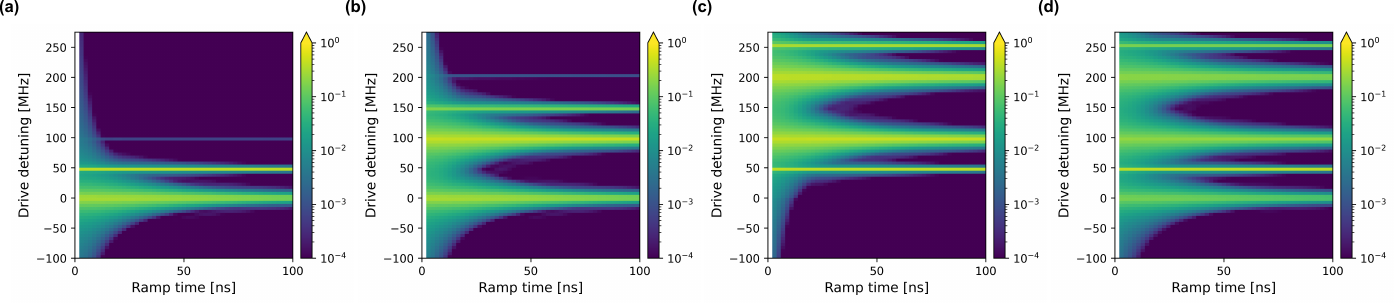}
    \caption{\label{fig:ancilla-init}\textbf{Ancilla-state dependence of non-adiabaticity.} Two-dimensional maps of the initial-state-dependent population infidelity over a $50\,\mathrm{ns}$ flat-top versus ramp time and drive detuning $\Delta_\mathrm{BS}$, using the same Hamiltonian and cosine ramp profile as in Fig.~\ref{fig:Fig3}b. The oscillator is initialized in an equal superposition of the first $N_c = 10$ Fock states with the transmon prepared in \textbf{(a)} $\lvert g\rangle$, \textbf{(b)} $\lvert e\rangle$, \textbf{(c)} $\lvert f\rangle$, and \textbf{(d)} $(\lvert g\rangle + \lvert f\rangle)/\sqrt{2}$. The active resonant swap poles shift depending on the occupied transmon level(s).}
\end{figure*}

To directly assess how faithfully the bosonic and ancilla populations map from the bare basis to the dressed basis without being obscured by dynamical phase errors, we define the population swapping error as a function of the ramp-up time $t_{\rm{r}}$,
\begin{equation}
\label{eq:pop-error}
    \varepsilon_{\mathrm{pop}}(t_{\rm{r}}) = \frac{1}{\lvert\mathcal{S}\rvert N_c} \sum_{j\in\mathcal{S}} \sum_{n=0}^{N_c-1} \left( 1 - \lvert \bra{n,j}_d \mathcal{U}(t_{\rm{r}},0) \ket{n,j}_b \rvert^2 \right),
\end{equation}
where $\ket{n,j}_b$ denotes the bare eigenstates before the ramp, $\ket{n,j}_d$ the dressed eigenstates after activation, and $\mathcal{U}(t_{\rm{r}}, 0)$ the propagator associated with activating the detuned-beamsplitter interaction. Here, $n \in \{0, \ldots, N_c-1\}$ indexes the oscillator Fock states and $j \in \mathcal{S}$ indexes the occupied transmon levels in the initial set $\mathcal{S}$ (when summed over all $N_q$ transmon levels, the normalization factor becomes $N_q \times N_c$, corresponding to $\varepsilon_{\mathrm{pop}}(t_{\rm{r}})$ in Fig.~\ref{fig:Fig3}(c)).

As in the main text, we assume a $gf$ qubit as the ancilla: $\lvert g\rangle$ and $\lvert f\rangle$ form the computational subspace, while $\lvert e\rangle$ is reserved for $T_1$ error detection so that a single-photon decay out of $\lvert f\rangle$ occupies a detectable flag level rather than inducing a computational error. 
We therefore evaluate the population error with the transmon initialized in the individual levels $\lvert g\rangle$, $\lvert e\rangle$, or $\lvert f\rangle$, as well as in the computational superposition $(\lvert g\rangle+\lvert f\rangle)/\sqrt{2}$.

To maintain the same conditions as in the main text, we use the same Hamiltonian and cosine ramp profile, and we evaluate the initial-state-dependent population infidelity (taking the maximum over a $50\,\mathrm{ns}$ flat-top). 
In our simulations, the oscillator is prepared in an equal superposition of the first $N_c=10$ Fock states while the transmon set $\mathcal{S}$ is varied, corresponding to the initial state
\begin{equation}
\label{eq:psi0-alln}
\lvert\psi_0(\mathcal{S})\rangle
=
\frac{1}{\sqrt{\lvert\mathcal{S}\rvert N_c}}
\sum_{j\in\mathcal{S}}
\sum_{n=0}^{N_c-1}
\lvert n,j\rangle_b.
\end{equation}
The resulting two-dimensional maps are shown for $\mathcal{S}=\{g\}$, $\{e\}$, $\{f\}$, and $\{g,f\}$ in Fig.~\ref{fig:ancilla-init}.

Decomposing the error in this manner directly reveals how each transmon level contributes to the multi-pole landscape observed in Fig.~\ref{fig:Fig3}(b). These poles correspond to resonant swap interactions between the ancilla and the bosonic mode, hence, we must operate sufficiently far from them. Unlike a two-level qubit, which poles are active depends on the occupied transmon level:
\begin{itemize}
    \item For $\lvert g\rangle$ [Fig.~\ref{fig:ancilla-init}(a)], the dominant features are the single-photon $g\leftrightarrow e$ resonance at $\Delta_\mathrm{BS}=0$ and the two-photon $g\leftrightarrow f$ process near $+50\,\mathrm{MHz}$.
    \item For $\lvert e\rangle$ [Fig.~\ref{fig:ancilla-init}(b)], the $T_1$-detection level, the map exhibits both the $g\leftrightarrow e$ pole at $\Delta_\mathrm{BS}=0$ and the $e\leftrightarrow f$ pole near $+100\,\mathrm{MHz}$.
    \item For $\lvert f\rangle$ [Fig.~\ref{fig:ancilla-init}(c)], the other computational state, the $e\leftrightarrow f$ pole near $+100\,\mathrm{MHz}$ and the two-photon $g\leftrightarrow f$ pole near $+50\,\mathrm{MHz}$ appear from the opposite side of the transitions; the next ladder transition, $f\leftrightarrow h$, lies near $+200\,\mathrm{MHz}$.
\end{itemize}
Away from these active poles, the error falls with ramp time in the same manner as in Fig.~\ref{fig:Fig3}(b), indicating that the two-level speed limit derived in Appendix~C still captures the behavior once the closest active pole is identified.

Finally, the $(\lvert g\rangle+\lvert f\rangle)/\sqrt{2}$ panel [Fig.~\ref{fig:ancilla-init}(d)] represents the initialization relevant to a $gf$-qubit protocol, in which a computational superposition is held while the dynamic $\chi$ is turned on. 
This map is the occupation-weighted combination of the $\lvert g\rangle$ and $\lvert f\rangle$ panels, where both families of poles appear, and no new resonance is introduced. The adiabatic condition for a $gf$ logical operation is therefore set by remaining sufficiently detuned from both the $\lvert g\rangle$ and $\lvert f\rangle$ poles, while the $\lvert e\rangle$ map governs the population error once a $T_1$-detected ancilla decay into $\lvert e\rangle$ has occurred.

\section{Numerical simulations of low-frequency dephasing}
\label[app]{app:low_freq_echo}
Many dephasing mechanisms in superconducting circuits exhibit a predominantly low-frequency noise profile, and therefore accurate noise modelling should take this into account. However, common methods of modelling decoherence such as the the Lindblad Master Equation method make a Markovian assumption of the environment, implicitly assuming white-noise dephasing. To remedy this, we implement an alternative monte-carlo based technique which we detail in the appendix below. At a high level, we generate noise trajectories whose serial behavior exhibits whatever noise spectrum we desire, then for each ``shot'' we couple one trajectory to $q^\dagger q$ to simulate frequency noise, and then finally we ensemble average over many noise trajectories.

\begin{figure}[h!]
    \centering\includegraphics[width=\columnwidth]{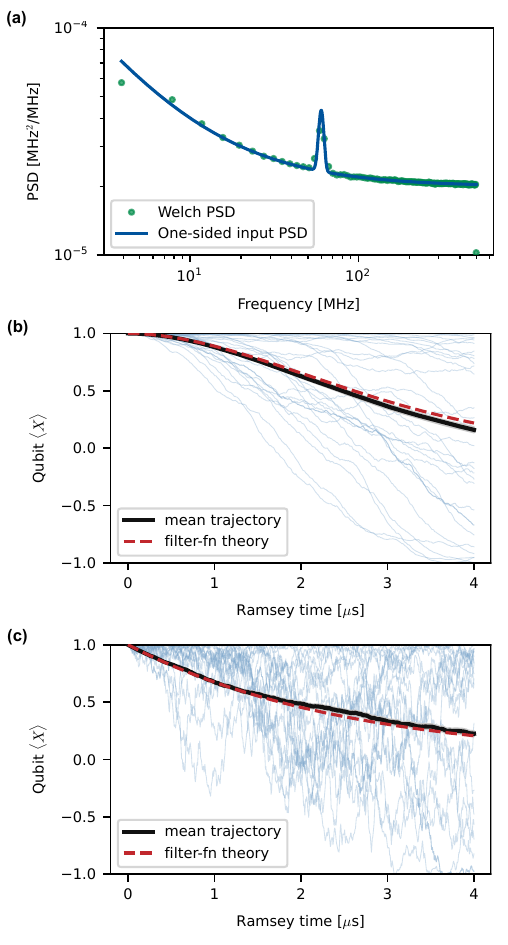}
    \caption{\textbf{Modeling low-frequency noise}. \textbf{(a)}, Example noise PSD, with a low-frequency frequency noise power of $S_{ff}[1 \textrm{ MHz}] = 10^{-4}\textrm{ MHz}$, white noise floor of $S_{ff} = 10^{-5}\textrm{ MHz}$, and a 60 Hz noise peak with power $10^{-5}\textrm{ MHz}$.
    \textbf{(b)}, Corresponding Qubit Ramsey decay of $\langle X\rangle$, with some individual trajectories plotted in light blue.
    \textbf{(c)}, Reference Ramsey decay with pure white (Markovian) noise of power $S_{ff} = 5\times10^{-3}\textrm{ MHz}$, and individual trajectories.
    }
    \label[fig]{fig:Noise_PSD}
\end{figure}

\begin{figure*}[ht]
    \centering\includegraphics[width=2.0\columnwidth]{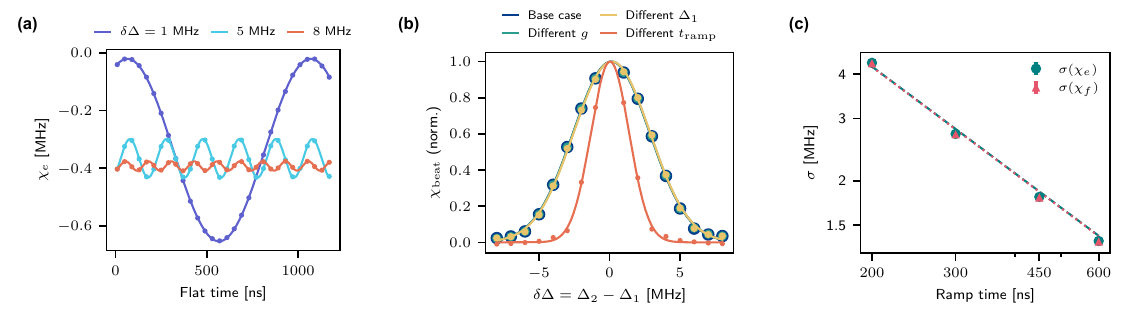}
    \caption{\label[fig]{fig:chi_beat_note}\textbf{Multi-tone interference effects. (a)},
    Instantaneous dispersive shift, extracted from the rate of change of phase accumulation of the states {$\ket{g0},\ket{g1},\ket{e0},\ket{e1}$} as a function of time. When the two drive tones are close in frequency, the dispersive shift oscillates at their relative detuning $\delta\Delta$. However, the amplitude of the oscillation decays with increasing $\delta\Delta$.
    \textbf{b}, 
    Fractional intensity of the $\chi(t)$ beat note as a function of relative detuning $\delta\Delta$. The beat note intensity decays in a gaussian manner, with Gaussian's linewidth being independent of drive strength $g$ and the absolute detuning from the nearest pole. However, the linewidth does depend on the dynamic-$\chi$ ramp time $T_{\text{ramp}}$.
    \textbf{c}, As $T_{\text{ramp}}$ is swept, the gaussian linewidth, and therefore the characteristic $\delta\Delta$ to observe interference effects, scales as $1/T_{\text{ramp}}$. This is the case both for the oscillating $\chi_e$ and for the oscillating $\chi_f$.}
\end{figure*}

To simulate more generic noise, we develop a workflow that takes a desired power spectral density as input and generates noise realizations consistent with that spectrum. 
Let \(\delta\omega(t)\) describe the instantaneous fluctuation of the ancilla angular frequency. Assuming this process is wide-sense stationary, we define its power spectral density as
\begin{align}
S_{\omega\omega}[f]
&= \lim_{T\to\infty}\frac{1}{T}
\left\langle\left|\delta\Omega_T[f]\right|^2\right\rangle, \\
\delta\Omega_T[f]
&= \int_0^T \delta\omega(t)e^{-i2\pi ft}\,dt,
\end{align}
where the brackets denote an ensemble average over noise realizations.
% \JWOG{not happy with notation}
Concretely, we sample the input PSD at evenly spaced frequencies $f_k = k/T_{\mathrm{exp}}$, from $1/T_{\mathrm{exp}}$ to $1/(2T_{\mathrm{samp}})$, where $T_{\mathrm{exp}}$ is the total duration of the generated noise record and $T_{\mathrm{samp}}$ is the time interval between samples.
We choose $T_{\mathrm{exp}} = N_{\mathrm{traj}}T_{\mathrm{traj}}$, so that the generated record can be divided into $N_{\mathrm{traj}}$ noise trajectories of duration $T_{\mathrm{traj}}$.
We then take the square root of the sampled PSD values to convert from spectral power to Fourier amplitude. 
If we were to perform an inverse FFT on this signal, it would not give us noise, as all spectral components are still purely real numbers, and therefore all in-phase with each other. 
To remedy this, we randomize the phase of each sample in frequency space before performing the iFFT to properly mimic noise, giving a time-domain signal of time $T_{exp}$. Finally, we cut this signal into snippets of time $T_{traj}$ to be used as individual noise trajectories to be coupled to $q^\dagger q$.

To validate this procedure, we both compare the spectrum of the generated noise with the input PSD and compare the resulting Ramsey decay with the prediction from filter-function theory. To show the versatile nature of this technique, we simulate $1/f$-type noise with a white noise floor, along with a singular peak at 60 Hz to mimic what may be seen in a lab. \cref{fig:Noise_PSD}a compares the input PSD with an estimate obtained from the generated noise using Welch's method \cite{Welch1967}, showing agreement across the resolved frequency range.
We then use these noise trajectories to simulate a Ramsey experiment, with individual realizations and their ensemble average shown in \cref{fig:Noise_PSD}b for the spectrum shown in \cref{fig:Noise_PSD}a, and for a white-noise spectrum in \cref{fig:Noise_PSD}c.
Within the Gaussian-noise approximation, the coherence envelope is given by $W(t)=e^{-\Xi(t)}$, where $\Xi(t)$ is obtained by integrating the product of the noise PSD and the sequence-dependent filter function,
\begin{equation}
\Xi(t)=\int_{1/T_{\mathrm{exp}}}^{1/(2T_{\mathrm{samp}})}
S_{\omega\omega}[f]F(f,t)\,df.
\end{equation}
Here, $S_{\omega\omega}[f]$ is the two-sided PSD of the transition angular-frequency fluctuation, and the equal contributions from positive and negative frequencies have been combined. For a Ramsey sequence, the filter function is
\begin{equation}
F_R(f,t)=t^2\left[\frac{\sin(\pi ft)}{\pi ft}\right]^2.
\end{equation}
The resulting coherence envelope agrees with the ensemble-averaged simulated Ramsey decay shown in both \cref{fig:Noise_PSD}b and \cref{fig:Noise_PSD}c, showing that this simulation technique both recovers the expected exponential decay from white noise dephasing while also allowing for more complex noise spectra as well.

\section{Multi-tone interference effects}
\label[app]{app:multi-tone_intereference}

In the main text, we assert that when turning on a multi-tone parametric beamsplitter interaction, the single-tone effective Hamiltonians simply add so long as the two tones are ``sufficiently detuned''. Here we back this assertion with numerics and explore its limits, revealing interference effects between two tones. Specifically, we show that this interference is suppressed when $|\delta\Delta| = |\Delta_{BS,2}-\Delta_{BS,1}| \gg 1/T_{\rm{ramp}}$, with a Gaussian roll-off in the amplitude of the interference. For the applications we propose, the two tones are spectrally separated by $\mathcal{O}(K)$ since they are near different poles, meaning that this multi-tone adiabatic condition is automatically satisfied so long as the single-tone adiabatic conditions are met.

To see why we might expect interference effects in the first place, note that the two tones enter through a single complex  amplitude
$g(t) = g_1(t) e^{i\Delta_1 t} + g_2(t) e^{i\Delta_2 t}$, and any drive-induced static quantity is proportional to an even power of this amplitude. 
For instance, the dispersive shift enters at leading order through
\begin{equation}
    |g(t)|^2 = g_1(t)^2 + g_2(t)^2 + 2 g_1(t) g_2(t) \cos(\delta\Delta\, t)
    \label{eq:two_tone_cross_term}
\end{equation}
where the squared terms reproduce the additive baseline
$\chi_1 + \chi_2$, but the cross term oscillates at exactly the beat note between the tones, $\delta\Delta$.
This picture, however, accounts only for the beat frequency: taken at face value it predicts a beat amplitude $\propto 2 g_1 g_2$, but with no clear prediction of how $\chi_{\rm beat}$ depends on $\delta\Delta$ itself. As is already visible in
\cref{fig:chi_beat_note}(a), and as we quantify below, the observed
$\chi_{\rm beat}$ instead collapses rapidly with tone separation.

\begin{figure}[ht]
    \centering\includegraphics[width=1.0\columnwidth]{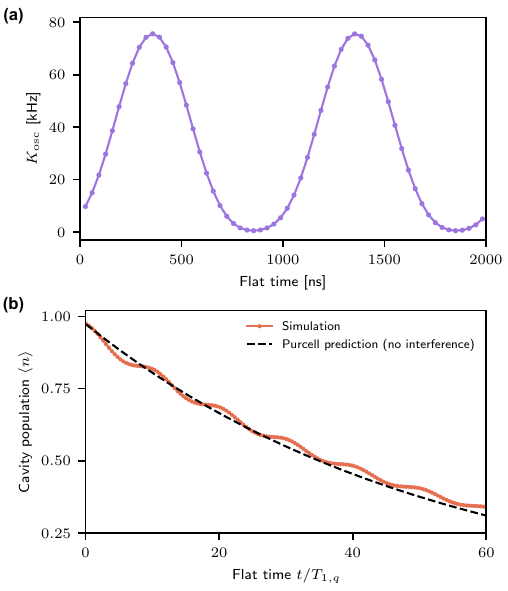}
    \caption{\textbf{Interference effects in the effective hybridization. (a),} 
    Instantaneous inherited oscillator Kerr as a function of time for a multi-tone drive. Again, we see oscillatory behavior at rate $\delta\Delta$.
    \textbf{ (b), } Oscillator population as a function of time in the presence of a multi-tone drive to a lossy ancilla. We observe an oscillation in the decay rate of the oscillator, showing that both coherent and incoherent phenomena show interference effects due to the modulated hybridization.
    }
    \label[fig]{fig:oscillating_Kerr_Purcell}
\end{figure}

Interestingly, the amplitude of the beat note $\chi_{\rm beat}$ drops off in a Gaussian manner as $\delta\Delta$ increases. The frequency scale $\sigma_{beat}$ that $\delta\Delta$ must be compared to appears to be independent of both the beamsplitter strengths and the general proximity to the nearest pole. However, $\sigma_{beat}$ is dependent upon the only remaining energy scale, the ramp time. We find that $\sigma_{beat}$ scales as $1/T_{\text{ramp}}$, implying that these interference effects arise from a lack of adiabaticity with respect to the beat note of the drives.

For the applications presented in this work, this adiabaticity condition is easily satisfied. When employing a multi-tone drive, the drive tones are placed on different sides of poles, meaning that any two tones are separated by $\mathcal{O}(K)$ in frequency, or at the very least separated by twice the distance to the nearest pole. Therefore, so long as each tone satisfies its adiabiticity condition with respect to its nearest pole, the multi-tone interference effects are automatically suppressed. 

Finally, we note that these interference effects are not specific to the dispersive interaction, but are symptomatic of beating in the underlying hybridization, which affects other coherent terms and inherited dissipation as well. 
This is the case as they all derive from the same modulated linear coupling. 
% \JWOG{needed?}
As illustrated in \cref{fig:oscillating_Kerr_Purcell}, both the inherited oscillator Kerr and oscillator decay oscillate at rate $\delta\Delta$. 
Indeed, the Kerr oscillates non-sinusoidally, consistent with being proportional to $|g(t)|^4$, or the square of \cref{eq:two_tone_cross_term}.
% \JWOG{not quite right, needs revision}
Together, these observations support a picture in which interference between the drive tones periodically strengthens and weakens the oscillator’s hybridizing with the nonlinear, lossy ancilla. 
While these interesting phenomena demonstrate a fascinating breakdown of the RWA, all applications presented in this paper operate in the limit where multi-tone adiabatic conditions are thoroughly satisfied, easing our analysis to the case where the effective Hamiltonians from individual tones effectively add. 

\section{Analytic $\chi$-matched echoed parity map}
\label[app]{app:analytic_gf_echo}
In the main text, we showed the power of applying optimal control theory to this bosonic control technique. Here, we detail an alternative analytic pulse sequence that combines dynamical decoupling techniques with a ``$\chi$-matched'' dispersive shift to achieve an ancilla-tolerant parity map that recovers a significant fraction of the OCT pulse performance. 

Naively, one may think to combine decay and dephasing resilience by activating positive and negative $\chi$-matched dispersive shifts for equal time, with an echo pulse in-between. However, because the $\ket{f}$-state population is affected by the echo but the $\ket{e}$-state population is not, the total phase accumulated when a decay event occurs depends on which half of the sequence the decay occured in. This which-path information ruins the ancilla-tolerance of the gate. 

\begin{figure}[ht]
    \centering\includegraphics[width=1.0\columnwidth]{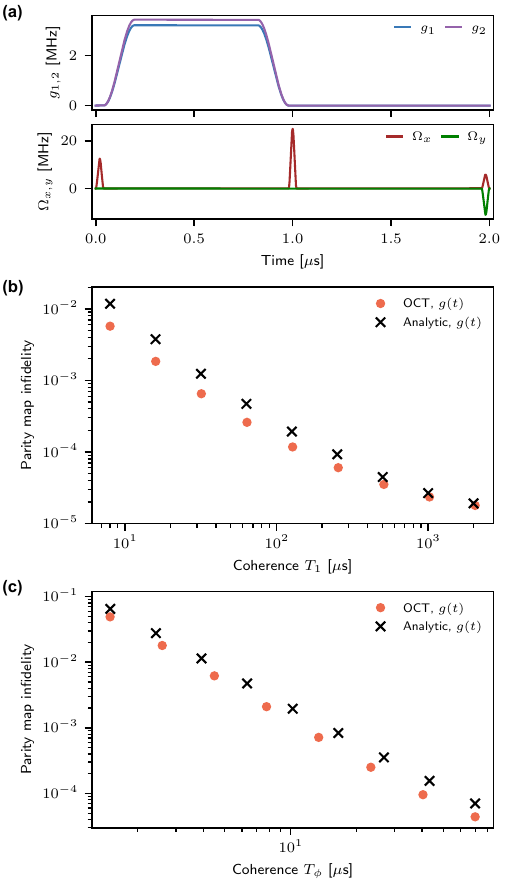}
    \caption{\textbf{Comparison of an analytic echoed g-f parity map to an optimized pulse. (a),} 
    Pulse sequence for an analytic parity map resilient to both ancilla decay and dephasing. The protocol takes a $T_2$ echo protocol in the $g-f$ manifold of the qubit and inserts a $\chi-$matched conditional phase accumulation in the first half of the sequence. 
    \textbf{ (b),(c) } Comparison of the simulated parity map infidelity for the OCT pulse shown in the main text against the analytic sequence shown in \textbf{(a)}, as a function of qubit $T_\phi$ and $T_1$.
    }
    \label[fig]{fig:analytic_vs_OCT}
\end{figure}

To remedy this, we can instead append an echo pulse and a fixed delay time to the $\chi$-matched parity map as shown in \cref{fig:analytic_vs_OCT}a. The protocol can be thought of from the ancilla's perspective as a $T_2$ echo sequence, where any phase accumulated equally in both halves will be cancelled, but now putting all conditional phase accumulation in the first half of the sequence. Additionally, because all $g(t)$ drives are off after the conditional phase is accumulated, the sequence maintains tolerance to ancilla decay. While this protocol is simple to calibrate and operate, the main drawback is that the total sequence time is doubled (for fixed $\chi$).

We then compare this analytic pulse sequence to the optimize pulse shown in the main text. We find that both for decay and dephasing, the analytic pulse sequence recovers most, but crucially not all, of the fidelity improvements achieved by the numerically-optimized pulse. This is understandable, as the total infidelity should scale roughly quadratically with sequence time. More generally, while here we have shown an analytic pulse sequence that achieves the desired goal of a parity map, optimal control techniques may well leverage this new bosonic control protocol to achieve significantly more abstract operations at high fidelity, such as a full bosonic QEC subroutine.

\section{Optimal control details}
\label[app]{app:oct}

We explore direct numeric optimization towards obtaining pulses enabled by the dynamic-$\chi$ paradigm that are robust to ancilla decay and dephasing. 
To enable the possibility of protection from ancilla decay, we utilize the $\ket{g},\ket{f}$ states of the ancilla as our qubit. We then reserve the $\ket{e}$ state as a flag state indicating that a jump has occurred. To enable protection from dephasing, we allow the optimizer direct control over drives on the $\ket{g},\ket{f}$ manifold of the ancilla. Despite these control knobs, it is non obvious that such a protected pulse should exist. For instance, once the ancilla decays, the $\ket{g},\ket{f}$ drives cease to act, which may result in the desired unitary not being completed. Because this decay can occur at any time during the gate operation, the desired unitary may not have been implemented. Nevertheless, we demonstrate below how such a pulse can be obtained with numerical optimization.

This optimization proceeds in multiple stages. In the first stage, we optimize the control pulses associated with the ``$\chi(t)$" Hamiltonian Eq.~\eqref{eq:H_oct_1} for both decay and dephasing protection. In the second stage, these pulses are translated into $g(t), \Delta$ drives as in Eq.~\eqref{eq:H_oct_2} and re-optimized. We now describe each stage in detail.

First, we optimize a parity measurement in the absence of decay or dephasing, using the cost function \cite{ReinholdThesis, AbdelhafezThesis}
\begin{align}
\label{eq:oct_C_infidelity}
C_{\rm infidelity} = \sum_{i}[1-\vert \langle \psi_{i}(T)\vert \psi_{i}^{\rm target}\rangle\vert^2].
\end{align}
The sum on $i$ is over the cardinal states of the Bloch spheres of the code and error spaces, with the ancilla always starting in the ground state. The states $\vert \psi_{i}^{\rm target}\rangle$ are the target final states, and $\vert \psi_{i}(T)\rangle$ are the states obtained from numerical integration at time $T$ (the gate time). This optimization yields initial seed pulses $\chi_{f}(t), I_{gf}(t)$ and $Q_{gf}(t)$ appearing in Eq.~\ref{eq:H_oct_1}. Notably, this does not give us a seed pulse yet for $\chi_{e}(t),$ because there is no mechanism yet for the ancilla to occupy the state $|e\rangle$.

Next, we use this pulse as a seed for an optimization that includes transmon decay. We additionally take as our initial guess $\chi_{e}(t)=\chi_{f}(t)$, taking inspiration from the known ``$\chi$-matching'' technique for protecting a bosonic mode against ancilla decay \cite{Reinhold2020} (note that some optimization rounds did obtain a $\chi$-matched pulse with only a random seed). 
Explicitly, the cost function we optimize is 
\begin{equation}
    \begin{split}
        C_{\rm infidelity}^{T_{1}} = \sum_{i}\bigg[1&-\Tr(P_{gf}\rho_{i}(T)P_{gf}\rho_{i}^{\rm target})\\
&-\Tr(P_{e}\rho_{i}(T)P_{e}\rho_{i}^{\rm jump})\bigg],
    \end{split}
\end{equation}
where $\rho_{i}(T)$ is now the numerically obtained density matrix at the gate time $T$. The density matrix is obtained using $\texttt{dynamiqs}$ \cite{dynamiqs} with either the \texttt{mesolve} or \texttt{jssesolve} (which we have contributed) methods: we have used both in the context of this optimization.
We have defined the projectors $P_{gf}=\vert g\rangle\langle g\vert+\vert f\rangle\langle f\vert, $ and $P_{e}=\vert e\rangle\langle e\vert$. These play the role of measurements of the ancilla, indicating either no ancilla decay has occurred (with the measurement result then used to infer whether the oscillator is in the code or error spaces) or that an ancilla decay has occurred and the parity measurement should be retried. The ideal final states in the case of a jump should thus be the 
initial bosonic mode states, together with transmon state $\vert e\rangle$ (without any acquired relative phases); for instance $\vert\psi_{0_{\rm L}}^{\rm jump}\rangle=\vert 0_{\rm L},e\rangle$ and $\vert\psi_{1_{\rm E}}^{\rm jump}\rangle=\vert 1_{\rm E},e\rangle$. These final states represent the ideal case where the jump has imparted no backaction on the bosonic mode state, and the parity measurement can be retried after one measures that the ancilla has decayed.
%measurement of the ancilla reveals a jump has taken place. 

Notice that the cost function $C_{\rm fidelity}^{T_{1}}$ naturally weighs the two contributions according to the probability that a jump occurs $p_{i}^{\rm jump} = \Tr{P_{e}\rho_{i}(T)}$. Indeed, observe that in the unitary case where no jumps in the ancilla occur, we obtain $C_{\rm infidelity}^{T_{1}}\rightarrow C_{\rm infidelity}$.

To now build in protection against low-frequency dephasing noise on the ancilla, we add an additional stochastic term to the Hamiltonian
\begin{align}
\label{eq:H_1_over_f}
H_{1/f}=\sum_{i=g,e,f}\eta_{i}(t)\ket{i}\bra{i},
\end{align}
where we generate the noisy signal $\eta_{i}(t)$ by applying a $1/f$ filter with a cutoff of 1 kHz to a randomly generated white-noise signal. We then batch over many instances of Eq.~\eqref{eq:H_1_over_f} and minimize the average infidelity. The cost function in this case (ignoring $T_{1}$) is
\begin{align}
C_{\rm infidelity}^{T_{\phi}} =\sum_{k}\sum_{i}
[1-\vert \langle \psi_{i}^{(k)}(T)\vert\psi_{i}^{\rm target}\rangle\vert^2],
\end{align}
where $k$ indexes the different $1/f$ noise instances. Towards obtaining simultaneous $T_{1}$ and $T_{\phi}$ protection, we use the $T_{1}$ optimized pulse as a seed for optimization of $C_{\rm infidelity}^{T_{\phi}}$.

One can additionally choose to optimize against $T_{1}$ and $T_{\phi}$ simultaneously, using the combined cost function
\begin{align}
\label{eq:cost_fun_all}
C_{\rm infidelity}^{T_{1}, T_{\phi}} =\sum_{k}\sum_{i}\bigg[1-&\Tr(P_{gf}\rho_{i}^{(k)}(T)P_{gf}\rho_{i}^{\rm target}) \\ \nonumber &-\Tr(P_{e}\rho_{i}^{(k)}(T)P_{e}\rho_{i}^{\rm jump})\bigg].
\end{align}
However, in practice we have not found that this step is necessary for achieving simultaneous protection and is, in a sense, ``overkill."

We now take the obtained $\chi_{e}(t), \chi_{f}(t)$ pulses and translate them into $g(t),\Delta$ pulses using results from \cref{app:PertTheory}.
We allow for four independent beamsplitter drives, with $\Delta_{1}/2\pi = -25$ MHz, $\Delta_{2}/2\pi = 75$ MHz, $\Delta_{3}/2\pi = 175$ MHz, and $\Delta_{4}/2\pi = 225$ MHz. These choices of detunings allow us to simultaneously drive near all relevant poles, and allow for the realization of nearly independent $\chi_{e}(t), \chi_{f}(t)$.
After constraining the four candidate detunings, we constrain a max change in $g_{BS}$ for each of the tones per unit time, automatically satisfying the constraints derived in \cref{app:speed_limit}. Then, starting from zero amplitude, we iteratively optimize drive strengths $g(t_{n+1})$ seeded by the drive strengths at prior time $t_n$. At each time point we optimize for both matching the dispersive shifts at time $t_{n+1}$, minimizing $\chi_e'$ and $\chi_f'$, and minimizing the total drive strength used. Together, this yields continuous $g(t)$ pulses for each detuning that simultaneously realize the desired $\chi(t)$ pulses while minimizing non-idealities and drive strength.

We now re-optimize the resulting pulse using Eq.~\eqref{eq:oct_C_infidelity}, to eliminate any non-idealities due to the translation step. This optimization now implicitly includes the effects of finite ramps, interference between the beamsplitter drives, etc. One can choose now to also re-optimize against decay and dephasing, however we have not found that to be necessary. 

\bibliography{Dynamic_chi_bibliography}% common bib file

@article{degraaf2025,
  title={A mid-circuit erasure check on a dual-rail cavity qubit using the joint-photon number-splitting regime of circuit QED},
  author={de Graaf, Stijn J and Xue, Sophia H and Chapman, Benjamin J and Teoh, James D and Tsunoda, Takahiro and Winkel, Patrick and Garmon, John W.O. and Chang, Kathleen M. and Frunzio, Luigi and Puri, Shruti and Schoelkopf, Robert J},
  journal={npj Quantum Information},
  doi = {10.1038/s41534-024-00944-4},
  url = {https://doi.org/10.1038/s41534-024-00944-4},
  volume={11},
  issue={1},
  year={2025}
}

@article{LiuISA2026,
  title = {Hybrid Oscillator-Qubit Quantum Processors: Instruction Set Architectures, Abstract Machine Models, and Applications},
  author = {Liu, Yuan and Singh, Shraddha and Smith, Kevin C. and Crane, Eleanor and Martyn, John M. and Eickbusch, Alec and Schuckert, Alexander and Li, Richard D. and Sinanan-Singh, Jasmine and Soley, Micheline B. and Tsunoda, Takahiro and Chuang, Isaac L. and Wiebe, Nathan and Girvin, Steven M.},
  journal = {PRX Quantum},
  volume = {7},
  issue = {1},
  pages = {010201},
  numpages = {166},
  year = {2026},
  month = {Jan},
  publisher = {American Physical Society},
  doi = {10.1103/4rf7-9tfx},
  url = {https://link.aps.org/doi/10.1103/4rf7-9tfx}
}

@article{tsunoda2023,
  title = {Error-Detectable Bosonic Entangling Gates with a Noisy Ancilla},
  author = {Tsunoda, Takahiro and Teoh, James D. and Kalfus, William D. and de Graaf, Stijn J. and Chapman, Benjamin J. and Curtis, Jacob C. and Thakur, Neel and Girvin, Steven M. and Schoelkopf, Robert J.},
  journal = {PRX Quantum},
  volume = {4},
  issue = {2},
  pages = {020354},
  numpages = {25},
  year = {2023},
  month = {Jun},
  publisher = {American Physical Society},
  doi = {10.1103/PRXQuantum.4.020354},
  url = {https://link.aps.org/doi/10.1103/PRXQuantum.4.020354}
}

@article{motzoi2009,
  title = {Simple Pulses for Elimination of Leakage in Weakly Nonlinear Qubits},
  author = {Motzoi, F. and Gambetta, J. M. and Rebentrost, P. and Wilhelm, F. K.},
  journal = {Phys. Rev. Lett.},
  volume = {103},
  issue = {11},
  pages = {110501},
  numpages = {4},
  year = {2009},
  month = {Sep},
  publisher = {American Physical Society},
  doi = {10.1103/PhysRevLett.103.110501},
  url = {https://link.aps.org/doi/10.1103/PhysRevLett.103.110501}
}

@article{theis2018,
  title = {Counteracting systems of diabaticities using DRAG controls: The status after 10 years},
  author = {Theis, L. S. and Motzoi, F. and Machnes, S. and Wilhelm, F. K.},
  journal = {Europhysics Letters},
  year = {2018},
  volume = {123},
  issue = {11},
  pages = {60001},
  publisher = {Europhysics Letters},
  doi = {10.1209/0295-5075/123/60001},
  url = {https://dx.doi.org/10.1209/0295-5075/123/60001}
}

@article{BeyondBreakeven2016,
	author = {Ofek, Nissim and Petrenko, Andrei and Heeres, Reinier and Reinhold, Philip and Leghtas, Zaki and Vlastakis, Brian and Liu, Yehan and Frunzio, Luigi and Girvin, S. M. and Jiang, L. and Mirrahimi, Mazyar and Devoret, M. H. and Schoelkopf, R. J.},
	da = {2016/08/01},
	doi = {10.1038/nature18949},
	id = {Ofek2016},
	isbn = {1476-4687},
	journal = {Nature},
	number = {7617},
	pages = {441--445},
	title = {Extending the lifetime of a quantum bit with error correction in superconducting circuits},
	ty = {JOUR},
	url = {https://doi.org/10.1038/nature18949},
	volume = {536},
	year = {2016}}

@article{GKPExperiment,
	author = {Campagne-Ibarcq, P. and Eickbusch, A. and Touzard, S. and Zalys-Geller, E. and Frattini, N. E. and Sivak, V. V. and Reinhold, P. and Puri, S. and Shankar, S. and Schoelkopf, R. J. and Frunzio, L. and Mirrahimi, M. and Devoret, M. H.},
	da = {2020/08/01},
	doi = {10.1038/s41586-020-2603-3},
	id = {Campagne-Ibarcq2020},
	isbn = {1476-4687},
	journal = {Nature},
	number = {7821},
	pages = {368--372},
	title = {Quantum error correction of a qubit encoded in grid states of an oscillator},
	ty = {JOUR},
	url = {https://doi.org/10.1038/s41586-020-2603-3},
	volume = {584},
	year = {2020}}

@article{sivak_real-time_2023,
	title = {Real-time quantum error correction beyond break-even},
	volume = {616},
	issn = {0028-0836, 1476-4687},
	url = {https://www.nature.com/articles/s41586-023-05782-6},
	doi = {10.1038/s41586-023-05782-6},
	number = {7955},
	urldate = {2024-07-25},
	journal = {Nature},
	author = {Sivak, V. V. and Eickbusch, A. and Royer, B. and Singh, S. and Tsioutsios, I. and Ganjam, S. and Miano, A. and Brock, B. L. and Ding, A. Z. and Frunzio, L. and Girvin, S. M. and Schoelkopf, R. J. and Devoret, M. H.},
	month = apr,
	year = {2023},
	pages = {50--55},
}

@article{Martinis2014,
  title = {Fast adiabatic qubit gates using only ${\ensuremath{\sigma}}_{z}$ control},
  author = {Martinis, John M. and Geller, Michael R.},
  journal = {Phys. Rev. A},
  volume = {90},
  issue = {2},
  pages = {022307},
  numpages = {9},
  year = {2014},
  month = {Aug},
  publisher = {American Physical Society},
  doi = {10.1103/PhysRevA.90.022307},
  url = {https://link.aps.org/doi/10.1103/PhysRevA.90.022307}
}

@unpublished{dynamiqs,
  title  = {Dynamiqs: an open-source Python library for GPU-accelerated and differentiable simulation of quantum systems},
  author = {Pierre Guilmin and Ronan Gautier and Adrien Bocquet and {\'{E}}lie Genois and Daniel K. Weiss},
  year   = {2024},
  url    = {https://github.com/dynamiqs/dynamiqs}
}

@unpublished{qontrol,
  title  = {qontrol: Quantum optimal control based on dynamiqs, diffrax and JAX},
  author = {Daniel K. Weiss},
  year   = {2024},
  url    = {https://github.com/dkweiss31/qontrol}
}

@article{Michael2016,
  title = {New Class of Quantum Error-Correcting Codes for a Bosonic Mode},
  author = {Michael, Marios H. and Silveri, Matti and Brierley, R. T. and Albert, Victor V. and Salmilehto, Juha and Jiang, Liang and Girvin, S. M.},
  journal = {Phys. Rev. X},
  volume = {6},
  issue = {3},
  pages = {031006},
  numpages = {26},
  year = {2016},
  publisher = {American Physical Society},
  doi = {10.1103/PhysRevX.6.031006},
  url = {https://link.aps.org/doi/10.1103/PhysRevX.6.031006}
}

@article{DiCarlo2009,
	author = {DiCarlo, L. and Chow, J. M. and Gambetta, J. M. and Bishop, Lev S. and Johnson, B. R. and Schuster, D. I. and Majer, J. and Blais, A. and Frunzio, L. and Girvin, S. M. and Schoelkopf, R. J.},
	date = {2009/07/01},
	doi = {10.1038/nature08121},
	id = {DiCarlo2009},
	isbn = {1476-4687},
	journal = {Nature},
	number = {7252},
	pages = {240--244},
	title = {Demonstration of two-qubit algorithms with a superconducting quantum processor},
	url = {https://doi.org/10.1038/nature08121},
	volume = {460},
	year = {2009}}

@article{yvonne_fluxTuning_2024,
	title = {On-demand transposition across light-matter interaction regimes in bosonic {cQED}},
	volume = {15},
	issn = {2041-1723},
	url = {https://www.nature.com/articles/s41467-024-50201-7},
	doi = {10.1038/s41467-024-50201-7},
	number = {1},
	urldate = {2025-01-14},
	journal = {Nature Communications},
	author = {Valadares, Fernando and Huang, Ni-Ni and Chu, Kyle Timothy Ng and Dorogov, Aleksandr and Chua, Weipin and Kong, Lingda and Song, Pengtao and Gao, Yvonne Y.},
	month = jul,
	year = {2024},
	pages = {5816},
}

@article{Swiadek2023,
  title = {Enhancing Dispersive Readout of Superconducting Qubits through Dynamic Control of the Dispersive Shift: Experiment and Theory},
  author = {Swiadek, Fran{\ifmmode \mbox{\c{c}}\else \c{c}\fi{}}ois and Shillito, Ross and Magnard, Paul and Remm, Ants and Hellings, Christoph and Lacroix, Nathan and Ficheux, Quentin and Zanuz, Dante Colao and Norris, Graham J. and Blais, Alexandre and Krinner, Sebastian and Wallraff, Andreas},
  journal = {PRX Quantum},
  volume = {5},
  issue = {4},
  pages = {040326},
  numpages = {14},
  year = {2024},
  month = {Nov},
  publisher = {American Physical Society},
  doi = {10.1103/PRXQuantum.5.040326},
  url = {https://link.aps.org/doi/10.1103/PRXQuantum.5.040326}
}

@phdthesis{AbdelhafezThesis,
  author       = {Mohamed Ragab Abdelhafez},
  title        = {Quantum Optimal Control Using Automatic Differentiation},
  school       = {University of Chicago},
  year         = {2019},
  type         = {Ph.D. Thesis},
  url          = {https://knowledge.uchicago.edu/record/2028},
}

@phdthesis{ReinholdThesis,
  author       = {Philip Reinhold},
  title        = {Controlling Error-Correctable Bosonic Qubits},
  school       = {Yale University},
  year         = {2019},
  type         = {Ph.D. Thesis},
}

@article{Reinhold2020,
	author = {Reinhold, Philip and Rosenblum, Serge and Ma, Wen-Long and Frunzio, Luigi and Jiang, Liang and Schoelkopf, Robert J.},
	date = {2020/08/01},
	doi = {10.1038/s41567-020-0931-8},
	id = {Reinhold2020},
	isbn = {1745-2481},
	journal = {Nature Physics},
	number = {8},
	pages = {822},
	title = {Error-corrected gates on an encoded qubit},
	url = {https://doi.org/10.1038/s41567-020-0931-8},
	volume = {16},
	year = {2020}}

@article{Yaxing_PRA_BilinearModeCoupling,
  title = {Engineering bilinear mode coupling in circuit QED: Theory and experiment},
  author = {Zhang, Yaxing and Lester, Brian J. and Gao, Yvonne Y. and Jiang, Liang and Schoelkopf, R. J. and Girvin, S. M.},
  journal = {Phys. Rev. A},
  volume = {99},
  issue = {1},
  pages = {012314},
  numpages = {39},
  year = {2019},
  month = {Jan},
  publisher = {American Physical Society},
  doi = {10.1103/PhysRevA.99.012314},
  url = {https://link.aps.org/doi/10.1103/PhysRevA.99.012314}
}

@article{circuitQEDReview,
  doi = {10.1103/revmodphys.93.025005},
  url = {https://doi.org/10.1103/revmodphys.93.025005},
  year = {2021},
  month = may,
  publisher = {American Physical Society ({APS})},
  volume = {93},
  number = {2},
  author = {Alexandre Blais and Arne L. Grimsmo and S.M. Girvin and Andreas Wallraff},
  title = {Circuit quantum electrodynamics},
  journal = {Reviews of Modern Physics}
}

@article{Krastanov2015,
  title = {Universal control of an oscillator with dispersive coupling to a qubit},
  author = {Krastanov, Stefan and Albert, Victor V. and Shen, Chao and Zou, Chang-Ling and Heeres, Reinier W. and Vlastakis, Brian and Schoelkopf, Robert J. and Jiang, Liang},
  journal = {Phys. Rev. A},
  volume = {92},
  issue = {4},
  pages = {040303},
  numpages = {5},
  year = {2015},
  month = {Oct},
  publisher = {American Physical Society},
  doi = {10.1103/PhysRevA.92.040303},
  url = {https://link.aps.org/doi/10.1103/PhysRevA.92.040303}
}

@misc{ChapmanSNAILBeamsplitterPaper,
  
  url = {https://arxiv.org/abs/2212.11929},
  
  author = {Chapman, Benjamin J. and de Graaf, Stijn J. and Xue, Sophia H. and Zhang, Yaxing and Teoh, James and Curtis, Jacob C. and Tsunoda, Takahiro and Eickbusch, Alec and Read, Alexander P. and Koottandavida, Akshay and Mundhada, Shantanu O. and Frunzio, Luigi and Devoret, M. H. and Girvin, S. M. and Schoelkopf, R. J.},
  
  title = {A high on-off ratio beamsplitter interaction for gates on bosonically encoded qubits},
  
  journal = {arXiv},
  
  year = {2022},
  
  copyright = {Creative Commons Attribution Non Commercial Share Alike 4.0 International}
}

@article{mirrahimi_cat_2014,
	title = {Dynamically protected cat-qubits: a new paradigm for universal quantum computation},
	volume = {16},
	issn = {1367-2630},
	shorttitle = {Dynamically protected cat-qubits},
	url = {https://iopscience.iop.org/article/10.1088/1367-2630/16/4/045014},
	doi = {10.1088/1367-2630/16/4/045014},
	
	number = {4},
	urldate = {2021-03-23},
	journal = {New Journal of Physics},
	author = {Mirrahimi, Mazyar and Leghtas, Zaki and Albert, Victor V and Touzard, Steven and Schoelkopf, Robert J and Jiang, Liang and Devoret, Michel H},
	month = apr,
	year = {2014},
	pages = {045014},
}

@article{Arne_rotating_2020,
  title = {Quantum Computing with Rotation-Symmetric Bosonic Codes},
  author = {Grimsmo, Arne L. and Combes, Joshua and Baragiola, Ben Q.},
  journal = {Phys. Rev. X},
  volume = {10},
  issue = {1},
  pages = {011058},
  numpages = {32},
  year = {2020},
  month = {Mar},
  publisher = {American Physical Society},
  doi = {10.1103/PhysRevX.10.011058},
  url = {https://link.aps.org/doi/10.1103/PhysRevX.10.011058}
}

@article{Hofheinz2009,
	author = {Hofheinz, Max and Wang, H. and Ansmann, M. and Bialczak, Radoslaw C. and Lucero, Erik and Neeley, M. and O'Connell, A. D. and Sank, D. and Wenner, J. and Martinis, John M. and Cleland, A. N.},
	date = {2009/05/01},
	doi = {10.1038/nature08005},
	id = {Hofheinz2009},
	isbn = {1476-4687},
	journal = {Nature},
	number = {7246},
	pages = {546--549},
	title = {Synthesizing arbitrary quantum states in a superconducting resonator},
	url = {https://doi.org/10.1038/nature08005},
	volume = {459},
	year = {2009}}

@article{SNAP_2015,
  title = {Cavity State Manipulation Using Photon-Number Selective Phase Gates},
  author = {Heeres, Reinier W. and Vlastakis, Brian and Holland, Eric and Krastanov, Stefan and Albert, Victor V. and Frunzio, Luigi and Jiang, Liang and Schoelkopf, Robert J.},
  journal = {Phys. Rev. Lett.},
  volume = {115},
  issue = {13},
  pages = {137002},
  numpages = {5},
  year = {2015},
  month = {Sep},
  publisher = {American Physical Society},
  doi = {10.1103/PhysRevLett.115.137002},
  url = {https://link.aps.org/doi/10.1103/PhysRevLett.115.137002}
}

@article{Rempe_cavity_QED,
author = {Gerhard Rempe},
title = {Atoms in an optical cavity: Quantum electrodynamics in confined space},
journal = {Contemporary Physics},
volume = {34},
number = {3},
pages = {119--129},
year = {1993},
publisher = {Taylor \& Francis},
doi = {10.1080/00107519308213809},

URL = { 
        https://doi.org/10.1080/00107519308213809
},
eprint = { 
        https://doi.org/10.1080/00107519308213809
}
}

@book{haroche2006exploring,
  title     = {Exploring the Quantum: Atoms, Cavities, and Photons},
  author    = {Haroche, Serge and Raimond, Jean-Michel},
  year      = {2006},
  publisher = {Oxford University Press},
  address   = {Oxford},
  isbn      = {978-0-19-850910-3},
  doi       = {10.1093/acprof:oso/9780198509103.001.0001}
}

@article{wallraff2004strong,
  title     = {Strong coupling of a single photon to a superconducting qubit using circuit quantum electrodynamics},
  author    = {Wallraff, Andreas and Schuster, D. I. and Blais, Alexandre and Frunzio, L. and Huang, R.-S. and Majer, J. and Kumar, S. and Girvin, S. M. and Schoelkopf, R. J.},
  journal   = {Nature},
  volume    = {431},
  number    = {7005},
  pages     = {162--167},
  year      = {2004},
  publisher = {Nature Publishing Group},
  doi       = {10.1038/nature02851}
}

@article{schoelkopf2008wiring,
  title     = {Wiring up quantum systems},
  author    = {Schoelkopf, R. J. and Girvin, S. M.},
  journal   = {Nature},
  volume    = {451},
  number    = {7179},
  pages     = {664--669},
  year      = {2008},
  publisher = {Nature Publishing Group},
  doi       = {10.1038/nature06179}
}

@misc{ParametricCQEDPaper,
  
  url = {https://arxiv.org/abs/2103.09277},
  
  author = {Noh, T. and Xiao, Z. and Cicak, K. and Jin, X. Y. and Doucet, E. and Teufel, J. and Aumentado, J. and Govia, L. C. G. and Ranzani, L. and Kamal, A. and Simmonds, R. W.},
  
  title = {Strong parametric dispersive shifts in a statically decoupled multi-qubit cavity QED system},
  
  publisher = {arXiv},
  
  year = {2021},
  
  copyright = {Creative Commons Attribution 4.0 International}
}

@article{ParametricCQEDTheory,
  title = {Perturbative Diagonalization for Time-Dependent Strong Interactions},
  author = {Xiao, Z. and Doucet, E. and Noh, T. and Ranzani, L. and Simmonds, R.W. and Govia, L.C.G. and Kamal, A.},
  journal = {Phys. Rev. Appl.},
  volume = {18},
  issue = {2},
  pages = {024009},
  numpages = {19},
  year = {2022},
  month = {Aug},
  publisher = {American Physical Society},
  doi = {10.1103/PhysRevApplied.18.024009},
  url = {https://link.aps.org/doi/10.1103/PhysRevApplied.18.024009}
}

@article{
PhilSergeFTDetection,
author = {S. Rosenblum  and P. Reinhold  and M. Mirrahimi  and Liang Jiang  and L. Frunzio  and R. J. Schoelkopf },
title = {Fault-tolerant detection of a quantum error},
journal = {Science},
volume = {361},
number = {6399},
pages = {266-270},
year = {2018},
doi = {10.1126/science.aat3996}}

@article{Akshay_chi_tuning,
  title = {Erasure Detection of a Dual-Rail Qubit Encoded in a Double-Post Superconducting Cavity},
  author = {Koottandavida, Akshay and Tsioutsios, Ioannis and Kargioti, Aikaterini and Smith, Cassady R. and Joshi, Vidul R. and Dai, Wei and Teoh, James D. and Curtis, Jacob C. and Frunzio, Luigi and Schoelkopf, Robert J. and Devoret, Michel H.},
  journal = {Phys. Rev. Lett.},
  volume = {132},
  issue = {18},
  pages = {180601},
  numpages = {7},
  year = {2024},
  month = {May},
  publisher = {American Physical Society},
  doi = {10.1103/PhysRevLett.132.180601},
  url = {https://link.aps.org/doi/10.1103/PhysRevLett.132.180601}
}

@article{JensTransmonPaper,
  title = {Charge-insensitive qubit design derived from the Cooper pair box},
  author = {Koch, Jens and Yu, Terri M. and Gambetta, Jay and Houck, A. A. and Schuster, D. I. and Majer, J. and Blais, Alexandre and Devoret, M. H. and Girvin, S. M. and Schoelkopf, R. J.},
  journal = {Phys. Rev. A},
  volume = {76},
  issue = {4},
  pages = {042319},
  numpages = {19},
  year = {2007},
  month = {Oct},
  publisher = {American Physical Society},
  doi = {10.1103/PhysRevA.76.042319},
  url = {https://link.aps.org/doi/10.1103/PhysRevA.76.042319}
}

@article{Fei2018,
  title = {Distinguishing Coherent and Thermal Photon Noise in a Circuit Quantum Electrodynamical System},
  author = {Yan, Fei and Campbell, Dan and Krantz, Philip and Kjaergaard, Morten and Kim, David and Yoder, Jonilyn L. and Hover, David and Sears, Adam and Kerman, Andrew J. and Orlando, Terry P. and Gustavsson, Simon and Oliver, William D.},
  journal = {Phys. Rev. Lett.},
  volume = {120},
  issue = {26},
  pages = {260504},
  numpages = {6},
  year = {2018},
  month = {Jun},
  publisher = {American Physical Society},
  doi = {10.1103/PhysRevLett.120.260504},
  url = {https://link.aps.org/doi/10.1103/PhysRevLett.120.260504}
}

@Article{SQUIDBeamsplitterPaper,
author={Lu, Yao
and Maiti, Aniket
and Garmon, John W. O.
and Ganjam, Suhas
and Zhang, Yaxing
and Claes, Jahan
and Frunzio, Luigi
and Girvin, Steven M.
and Schoelkopf, Robert J.},
title={High-fidelity parametric beamsplitting with a parity-protected converter},
journal={Nature Communications},
year={2023},
month={Sep},
day={18},
volume={14},
number={1},
pages={5767},
issn={2041-1723},
doi={10.1038/s41467-023-41104-0},
url={https://doi.org/10.1038/s41467-023-41104-0}
}

@article{Brune1992CatState,
  title = {Manipulation of photons in a cavity by dispersive atom-field coupling: Quantum-nondemolition measurements and generation of ``Schr\"odinger cat'' states},
  author = {Brune, M. and Haroche, S. and Raimond, J. M. and Davidovich, L. and Zagury, N.},
  journal = {Phys. Rev. A},
  volume = {45},
  issue = {7},
  pages = {5193--5214},
  numpages = {0},
  year = {1992},
  month = {Apr},
  publisher = {American Physical Society},
  doi = {10.1103/PhysRevA.45.5193},
  url = {https://link.aps.org/doi/10.1103/PhysRevA.45.5193}
}

@article{Bertet2002Fock1Wigner,
  title = {Direct Measurement of the Wigner Function of a One-Photon Fock State in a Cavity},
  author = {Bertet, P. and Auffeves, A. and Maioli, P. and Osnaghi, S. and Meunier, T. and Brune, M. and Raimond, J. M. and Haroche, S.},
  journal = {Phys. Rev. Lett.},
  volume = {89},
  issue = {20},
  pages = {200402},
  numpages = {4},
  year = {2002},
  month = {Oct},
  publisher = {American Physical Society},
  doi = {10.1103/PhysRevLett.89.200402},
  url = {https://link.aps.org/doi/10.1103/PhysRevLett.89.200402}
}

@article{
Wang2016CatTwoBoxes,
author = {Chen Wang  and Yvonne Y. Gao  and Philip Reinhold  and R. W. Heeres  and Nissim Ofek  and Kevin Chou  and Christopher Axline  and Matthew Reagor  and Jacob Blumoff  and K. M. Sliwa  and L. Frunzio  and S. M. Girvin  and Liang Jiang  and M. Mirrahimi  and M. H. Devoret  and R. J. Schoelkopf },
title = {A Schrödinger cat living in two boxes},
journal = {Science},
volume = {352},
number = {6289},
pages = {1087-1091},
year = {2016},
doi = {10.1126/science.aaf2941},
URL = {https://www.science.org/doi/abs/10.1126/science.aaf2941},
eprint = {https://www.science.org/doi/pdf/10.1126/science.aaf2941}}

@article{Schuster2007PhotonNumberSplitting,
	author = {Schuster, D. I. and Houck, A. A. and Schreier, J. A. and Wallraff, A. and Gambetta, J. M. and Blais, A. and Frunzio, L. and Majer, J. and Johnson, B. and Devoret, M. H. and Girvin, S. M. and Schoelkopf, R. J.},
	journal = {Nature},
	number = {7127},
	pages = {515--518},
	title = {Resolving photon number states in a superconducting circuit},
	volume = {445},
	year = {2007}}

@article{Law1996ResonantControl,
  title = {Arbitrary Control of a Quantum Electromagnetic Field},
  author = {Law, C. K. and Eberly, J. H.},
  journal = {Phys. Rev. Lett.},
  volume = {76},
  issue = {7},
  pages = {1055--1058},
  numpages = {0},
  year = {1996},
  month = {Feb},
  publisher = {American Physical Society},
  doi = {10.1103/PhysRevLett.76.1055},
  url = {https://link.aps.org/doi/10.1103/PhysRevLett.76.1055}
}

@article{Reagor2016QuantumMemory,
  title = {Quantum memory with millisecond coherence in circuit QED},
  author = {Reagor, Matthew and Pfaff, Wolfgang and Axline, Christopher and Heeres, Reinier W. and Ofek, Nissim and Sliwa, Katrina and Holland, Eric and Wang, Chen and Blumoff, Jacob and Chou, Kevin and Hatridge, Michael J. and Frunzio, Luigi and Devoret, Michel H. and Jiang, Liang and Schoelkopf, Robert J.},
  journal = {Phys. Rev. B},
  volume = {94},
  issue = {1},
  pages = {014506},
  numpages = {8},
  year = {2016},
  month = {Jul},
  publisher = {American Physical Society},
  doi = {10.1103/PhysRevB.94.014506},
  url = {https://link.aps.org/doi/10.1103/PhysRevB.94.014506}
}

@article{Zhu2013Guanyu4thOrder,
  title = {Circuit QED with fluxonium qubits: Theory of the dispersive regime},
  author = {Zhu, Guanyu and Ferguson, David G. and Manucharyan, Vladimir E. and Koch, Jens},
  journal = {Phys. Rev. B},
  volume = {87},
  issue = {2},
  pages = {024510},
  numpages = {16},
  year = {2013},
  month = {Jan},
  publisher = {American Physical Society},
  doi = {10.1103/PhysRevB.87.024510},
  url = {https://link.aps.org/doi/10.1103/PhysRevB.87.024510}
}

@article{Iivari2024,
  title = {Strategies and Trade-Offs for Controllability and Memory Time of Ultra-High-Quality Microwave Cavities in Circuit Quantum Electrodynamics},
  author = {Pietik\"ainen, Iivari and Cernot\'{\i}k, O. and Eickbusch, Alec and Maiti, Aniket and Garmon, John W.O. and Filip, Radim and Girvin, Steven M.},
  journal = {PRX Quantum},
  volume = {5},
  issue = {4},
  pages = {040307},
  numpages = {30},
  year = {2024},
  month = {Oct},
  publisher = {American Physical Society},
  doi = {10.1103/PRXQuantum.5.040307},
  url = {https://link.aps.org/doi/10.1103/PRXQuantum.5.040307}
}

@article{MaitiLINCTheory2025,
  title = {Linear quantum coupler for clean bosonic control},
  author = {Maiti, Aniket and Garmon, John W. O. and Lu, Yao and Miano, Alessandro and Frunzio, Luigi and Schoelkopf, Robert J.},
  journal = {PRX Quantum},
  pages = {--},
  year = {2025},
  month = {Jun},
  publisher = {American Physical Society},
  doi = {10.1103/7rnm-rxhh},
  url = {https://link.aps.org/doi/10.1103/7rnm-rxhh}
}

@article{Purcell1946,
  title = {Resonance Absorption by Nuclear Magnetic Moments in a Solid},
  author = {Purcell, E. M. and Torrey, H. C. and Pound, R. V.},
  journal = {Phys. Rev.},
  volume = {69},
  issue = {1-2},
  pages = {37--38},
  numpages = {0},
  year = {1946},
  month = {Jan},
  publisher = {American Physical Society},
  doi = {10.1103/PhysRev.69.37},
  url = {https://link.aps.org/doi/10.1103/PhysRev.69.37}
}

@article{Huang2025fastsidebandcontrolweakly,
  title = {Fast Sideband Control of a Multimode Cavity Memory with Weak Dispersive Coupling to a Transmon},
  author = {Huang, Jordan and DiNapoli, Thomas J. and Rockwood, Gavin and Yuan, Ming and Narasimhan, Prathyankara and Gupta, Eesh and Bal, Mustafa and Crisa, Francesco and Garattoni, Sabrina and Lu, Yao and Jiang, Liang and Chakram, Srivatsan},
  journal = {Phys. Rev. X},
  volume = {16},
  issue = {1},
  pages = {011058},
  numpages = {32},
  year = {2026},
  month = {Mar},
  publisher = {American Physical Society},
  doi = {10.1103/t4cv-y398},
  url = {https://link.aps.org/doi/10.1103/t4cv-y398}
}

@misc{huang2026JCcontrol,
      title={Universal Jaynes-Cummings Control of an Oscillator}, 
      author={Jordan Huang and Ethan Kasaba and Thomas J. DiNapoli and Tanay Roy and Srivatsan Chakram},
      year={2026},
      eprint={2605.18658},
      archivePrefix={arXiv},
      primaryClass={quant-ph},
      url={https://arxiv.org/abs/2605.18658}, 
}

@article{Jin2024ParametricCZNIST,
  title = {Superconducting architecture demonstrating fast, tunable high-fidelity CZ gates with parametric control of ZZ coupling},
  author = {Jin, X.Y. and Parrott, Z. and Cicak, K. and Kotler, S. and Lecocq, F. and Teufel, J. and Aumentado, J. and Kapit, E. and Simmonds, R.W.},
  journal = {Phys. Rev. Appl.},
  volume = {24},
  issue = {6},
  pages = {064026},
  numpages = {39},
  year = {2025},
  month = {Dec},
  publisher = {American Physical Society},
  doi = {10.1103/kmls-lgp5},
  url = {https://link.aps.org/doi/10.1103/kmls-lgp5}
}

@article{Bylander2011,
	author = {Bylander, Jonas and Gustavsson, Simon and Yan, Fei and Yoshihara, Fumiki and Harrabi, Khalil and Fitch, George and Cory, David G. and Nakamura, Yasunobu and Tsai, Jaw-Shen and Oliver, William D.},
	journal = {Nature Physics},
	number = {7},
	pages = {565--570},
	title = {Noise spectroscopy through dynamical decoupling with a superconducting flux qubit},
	volume = {7},
	year = {2011}}

@article{FeiYan2013,
	author = {Yan, Fei and Gustavsson, Simon and Bylander, Jonas and Jin, Xiaoyue and Yoshihara, Fumiki and Cory, David G. and Nakamura, Yasunobu and Orlando, Terry P. and Oliver, William D.},
	journal = {Nature Communications},
	number = {1},
	pages = {2337},
	title = {Rotating-frame relaxation as a noise spectrum analyser of a superconducting qubit undergoing driven evolution},
	volume = {4},
	year = {2013}}

@ARTICLE{Welch1967,
  author={Welch, P.},
  journal={IEEE Transactions on Audio and Electroacoustics}, 
  title={The use of fast Fourier transform for the estimation of power spectra: A method based on time averaging over short, modified periodograms}, 
  year={1967},
  volume={15},
  number={2},
  pages={70-73},
  doi={10.1109/TAU.1967.1161901}}

@article{WalraffDispersiveReadout2005,
  title = {Approaching Unit Visibility for Control of a Superconducting Qubit with Dispersive Readout},
  author = {Wallraff, A. and Schuster, D. I. and Blais, A. and Frunzio, L. and Majer, J. and Devoret, M. H. and Girvin, S. M. and Schoelkopf, R. J.},
  journal = {Phys. Rev. Lett.},
  volume = {95},
  issue = {6},
  pages = {060501},
  numpages = {4},
  year = {2005},
  month = {Aug},
  publisher = {American Physical Society},
  doi = {10.1103/PhysRevLett.95.060501},
  url = {https://link.aps.org/doi/10.1103/PhysRevLett.95.060501}
}

@article{WalraffDispersiveReadout2017,
  title = {Rapid High-Fidelity Single-Shot Dispersive Readout of Superconducting Qubits},
  author = {Walter, T. and Kurpiers, P. and Gasparinetti, S. and Magnard, P. and Poto{\ifmmode \check{c}\else \v{c}\fi{}}nik, A. and Salath{\'e}, Y. and Pechal, M. and Mondal, M. and Oppliger, M. and Eichler, C. and Wallraff, A.},
  journal = {Phys. Rev. Appl.},
  volume = {7},
  issue = {5},
  pages = {054020},
  numpages = {11},
  year = {2017},
  month = {May},
  publisher = {American Physical Society},
  doi = {10.1103/PhysRevApplied.7.054020},
  url = {https://link.aps.org/doi/10.1103/PhysRevApplied.7.054020}
}

@article{Majer2007,
	author = {Majer, J. and Chow, J. M. and Gambetta, J. M. and Koch, Jens and Johnson, B. R. and Schreier, J. A. and Frunzio, L. and Schuster, D. I. and Houck, A. A. and Wallraff, A. and Blais, A. and Devoret, M. H. and Girvin, S. M. and Schoelkopf, R. J.},
	date = {2007/09/01},
	doi = {10.1038/nature06184},
	id = {Majer2007},
	isbn = {1476-4687},
	journal = {Nature},
	number = {7161},
	pages = {443--447},
	title = {Coupling superconducting qubits via a cavity bus},
	url = {https://doi.org/10.1038/nature06184},
	volume = {449},
	year = {2007}}

@article{YaxingLeakageCancellation2025,
  title = {Active Leakage Cancellation in Single Qubit Gates},
  author = {Chiaro, Ben and Zhang, Yaxing},
  journal = {Phys. Rev. Lett.},
  volume = {135},
  issue = {13},
  pages = {130601},
  numpages = {8},
  year = {2025},
  month = {Sep},
  publisher = {American Physical Society},
  doi = {10.1103/4kz9-w97h},
  url = {https://link.aps.org/doi/10.1103/4kz9-w97h}
}

@article{FeiYanCouplerTheory2018,
  title = {Tunable Coupling Scheme for Implementing High-Fidelity Two-Qubit Gates},
  author = {Yan, Fei and Krantz, Philip and Sung, Youngkyu and Kjaergaard, Morten and Campbell, Daniel L. and Orlando, Terry P. and Gustavsson, Simon and Oliver, William D.},
  journal = {Phys. Rev. Appl.},
  volume = {10},
  issue = {5},
  pages = {054062},
  numpages = {9},
  year = {2018},
  month = {Nov},
  publisher = {American Physical Society},
  doi = {10.1103/PhysRevApplied.10.054062},
  url = {https://link.aps.org/doi/10.1103/PhysRevApplied.10.054062}
}

@article{SungTunableCouplerExperiment2021,
  title = {Realization of High-Fidelity CZ and $ZZ$-Free iSWAP Gates with a Tunable Coupler},
  author = {Sung, Youngkyu and Ding, Leon and Braum{\"u}ller, Jochen and Veps{\"a}l{\"a}inen, Antti and Kannan, Bharath and Kjaergaard, Morten and Greene, Ami and Samach, Gabriel O. and McNally, Chris and Kim, David and Melville, Alexander and Niedzielski, Bethany M. and Schwartz, Mollie E. and Yoder, Jonilyn L. and Orlando, Terry P. and Gustavsson, Simon and Oliver, William D.},
  journal = {Phys. Rev. X},
  volume = {11},
  issue = {2},
  pages = {021058},
  numpages = {32},
  year = {2021},
  month = {Jun},
  publisher = {American Physical Society},
  doi = {10.1103/PhysRevX.11.021058},
  url = {https://link.aps.org/doi/10.1103/PhysRevX.11.021058}
}

@article{PaikRIPGate2016,
  title = {Experimental Demonstration of a Resonator-Induced Phase Gate in a Multiqubit Circuit-QED System},
  author = {Paik, Hanhee and Mezzacapo, A. and Sandberg, Martin and McClure, D. T. and Abdo, B. and C{\'o}rcoles, A. D. and Dial, O. and Bogorin, D. F. and Plourde, B. L. T. and Steffen, M. and Cross, A. W. and Gambetta, J. M. and Chow, Jerry M.},
  journal = {Phys. Rev. Lett.},
  volume = {117},
  issue = {25},
  pages = {250502},
  numpages = {5},
  year = {2016},
  month = {Dec},
  publisher = {American Physical Society},
  doi = {10.1103/PhysRevLett.117.250502},
  url = {https://link.aps.org/doi/10.1103/PhysRevLett.117.250502}
}

@article{VlastakisBigCat2013,
	author = {Brian Vlastakis and Gerhard Kirchmair and Zaki Leghtas and Simon E. Nigg and Luigi Frunzio and S. M. Girvin and Mazyar Mirrahimi and M. H. Devoret and R. J. Schoelkopf},
	doi = {10.1126/science.1243289},
	eprint = {https://www.science.org/doi/pdf/10.1126/science.1243289},
	journal = {Science},
	number = {6158},
	pages = {607-610},
	title = {Deterministically Encoding Quantum Information Using 100-Photon Schr{\"o}dinger Cat States},
	url = {https://www.science.org/doi/abs/10.1126/science.1243289},
	volume = {342},
	year = {2013}}

@article{WeiSumeruReadout,
  title = {Benchmarking the Readout of a Superconducting Qubit for Repeated Measurements},
  author = {Hazra, S. and Dai, W. and Connolly, T. and Kurilovich, P. D. and Wang, Z. and Frunzio, L. and Devoret, M. H.},
  journal = {Phys. Rev. Lett.},
  volume = {134},
  issue = {10},
  pages = {100601},
  numpages = {6},
  year = {2025},
  month = {Mar},
  publisher = {American Physical Society},
  doi = {10.1103/PhysRevLett.134.100601},
  url = {https://link.aps.org/doi/10.1103/PhysRevLett.134.100601}
}

@article{KapitErrorTransparency2018,
  title = {Error-Transparent Quantum Gates for Small Logical Qubit Architectures},
  author = {Kapit, Eliot},
  journal = {Phys. Rev. Lett.},
  volume = {120},
  issue = {5},
  pages = {050503},
  numpages = {5},
  year = {2018},
  month = {Feb},
  publisher = {American Physical Society},
  doi = {10.1103/PhysRevLett.120.050503},
  url = {https://link.aps.org/doi/10.1103/PhysRevLett.120.050503}
}

@article{KeNieFluxoniumOscillator2026,
  title = {Fluxonium as a Control Qubit for Bosonic Quantum Information},
  author = {Nie, Ke and Bradford, J. Nofear and Mandal, Supriya and Bista, Aayam and Pfaff, Wolfgang and Kou, Angela},
  journal = {PRX Quantum},
  volume = {7},
  issue = {1},
  pages = {010357},
  numpages = {20},
  year = {2026},
  month = {Mar},
  publisher = {American Physical Society},
  doi = {10.1103/dgwk-w3jj},
  url = {https://link.aps.org/doi/10.1103/dgwk-w3jj}
}

@misc{ShoumikShantanuFluxoniumGKP2026Arxiv,
      title={Bosonic Error Correction with Fluxonium}, 
      author={Shantanu R. Jha and Shoumik D. Chowdhury and Gabriele Rolleri and Anaida Ali and Lev-Arcady Sellem and Réouven Assouly and David Pahl and Lukas Pahl and Junyoung An and Farid Hassani and Hung-Yu Tsao and Chia-Chin Tsai and Aranya Goswami and Gabriel D. Cutter and Jeremie Boudreault and Jeffrey M. Gertler and Michael A. Gingras and Bethany M. Niedzielski and Jeffrey M. Knecht and Mollie E. Schwartz and Kyle Serniak and Jeffrey A. Grover and Baptiste Royer and Max Hays and William D. Oliver},
      year={2026},
      eprint={2609.26880},
      archivePrefix={arXiv},
      primaryClass={quant-ph},
      url={https://arxiv.org/abs/2609.26880}, 
}

\end{document}